%% file: dichlong.tex
\documentclass{elsarticle} 

\usepackage{times}
\usepackage{amsmath}
\usepackage{bbm}
\usepackage[latin1]{inputenc}
\usepackage{latexsym}
\usepackage{epsfig}
\usepackage{amssymb}
\usepackage{graphicx}
\usepackage{colortbl}
\usepackage{color}
\usepackage{multirow}

\usepackage{comment}
\usepackage{booktabs}

\usepackage{rotating}
\usepackage[pagebackref]{hyperref}

\newtheorem{theorem}{Theorem}
\newtheorem{lemma}{Lemma}

\newtheorem{proposition}{Proposition}

\newproof{proof}{Proof}

\DeclareMathOperator{\fixed}{fixed}
\DeclareMathOperator{\sol}{sol}

\DeclareMathOperator{\al}{\alpha}

\begin{document}
\title{Towards a Dichotomy  for the Possible Winner Problem in Elections Based on Scoring Rules\tnoteref{t1}} 

\tnotetext[t1]{A preliminary version of this work appeared in \textit{Proceedings of the 34th International Symposium on Mathematical Foundations of Computer Science (MFCS'09), Novy Smokovec, Slovakia, August 2009, volume 5734 in Lecture Notes in Computer Science, pages 124--136}.}

\author[jena]{Nadja~Betzler\corref{cor}\fnref{fn1}}
\ead{nadja.betzler@uni-jena.de}

\author[tue]{Britta~Dorn}
\ead{bdorn@informatik.uni-tuebingen.de}

\cortext[cor]{Corresponding author, phone: +49 3641 9 46325, fax: +49 3641 9 46322  }

\fntext[fn1]{Supported by the DFG, project PAWS, NI~369/10.}  

\address[jena]{Institut f\"ur Informatik, Friedrich-Schiller-Universit\"at Jena, Ernst-Abbe-Platz 2, D-07743 Jena, Germany.}

\address[tue]{Wilhelm-Schickard-Institut f\"ur Informatik, Universit\"at T\"ubingen, Sand 13, D-72076 T\"ubingen, Germany.}


\begin{abstract} 
To make a joint decision, agents (or voters) are often required to
provide their preferences as linear orders. To determine a winner, the
given linear orders can be aggregated according to a voting
protocol. However, in realistic settings, the voters may often only
provide partial orders. This directly leads to the \textsc{Possible
Winner} problem that asks, given a set of partial votes, whether a
distinguished candidate can still become a winner. In this work, we
consider the computational complexity of \textsc{Possible Winner} for
the broad class of voting protocols defined by scoring rules. A
scoring rule provides a score value for every position which a
candidate can have in a linear order. Prominent examples include
plurality, $k$-approval, and Borda. Generalizing previous NP-hardness
results for some special cases, we settle the computational complexity
for all but one scoring rule. More precisely, for an unbounded number
of candidates and unweighted voters, we show that \textsc{Possible
Winner} is NP-complete for all pure scoring rules except plurality,
veto, and the scoring rule defined by the scoring
vector~$(2,1,\dots,1,0)$, while it is solvable in polynomial time for
plurality and veto.
\end{abstract}

\maketitle

\section{Introduction}
Voting scenarios arise whenever the preferences of different parties
({\it voters}) have to be aggregated to form a joint decision. This is
what happens in political elections, group decisions, web site
rankings, or multiagent systems. Often, the voting process is executed
in the following way: each voter provides his preference as a ranking
(linear order) of all the possible alternatives ({\it
candidates}). Given these rankings as an input, a {\it voting rule}
produces a subset of the candidates ({\it winners}) as an output.
However, in realistic settings,  the voters may often 
only provide partial orders (or partial votes) instead of linear ones: For example, it might be 
impossible for the voters to provide a complete preference list
because the set of candidates is too large, as it is the case for web page ranking. In addition, not all voters might
have given their preferences yet during the aggregation process, or
new candidates might be introduced after some voters already have
given their rankings. Moreover, one often has to deal with partial
votes due to incomparabilities: for some voters it might not be
possible to compare two candidates or certain groups of candidates, be
it because of lack of information or due to personal reasons.  Hence,
the study of partial voting profiles is  natural and essential. One question that immediately comes to mind is whether any
information on a possible outcome of the voting process can be given
in the case of incomplete votes.
More specifically, in this paper, we study the {\sc Possible Winner} problem: Given a partial order for each of the voters, can a distinguished candidate~$c$ win for at least one extension of the partial orders into linear ones?

Of course, the answer to this question depends on the voting rule that is used. In this work, we will stick to the broad class of  {\it scoring rules}. 
A scoring rule provides a score value for every position that a candidate can take within a linear order, given as a {\it scoring vector} of length~$m$ in the case of~$m$ candidates. The scores of the candidates are then added over all votes and the candidates with the highest score win.
Famous examples  are Borda, defined by the scoring vectors~$(m-1, m-2, \dots, 0 )$  and
$k$-approval, defined by~$(1,\dots,1,0, \dots ,0)$ starting with $k$ ones. Two
relevant special cases of $k$-approval are plurality, defined by~$(1,0,\dots, 0)$,
and veto, defined by~$(1, \dots, 1,0)$.
Typically, $k$-approval can be used in  political elections whenever the voters can express their preference for~$k$ candidates within the set of all candidates. Another example is the Formula 1 scoring, which until the year~2009 used the scoring rule defined by the vector~$(10,8,6,5,4,3,2,1,0,\dots, 0)$ and since~2010 
 uses~$(25,18,15,12,10,8,6,4,2,1,0,\dots, 0)$. 

The study of the computational complexity of voting problems is an active area of research (see the surveys~\cite{CELM07,FHHR09}). 
The \textsc{Possible Winner} problem was introduced by Konczak and
Lang~\cite{KL05} and has been further investigated since then for many
types of voting systems~\cite{BHN09,LPRVW07,PRVW07,Wal07,XC08}. Note
that the  related  \textsc{Necessary Winner} problem 
(Given a set of partial orders, does a distinguished candidate~$c$ win for every extension of the partial orders into linear ones?) 
can be
solved in polynomial time for all scoring rules~\cite{XC08}.

A prominent special case of \textsc{Possible Winner} is \textsc{Manipulation} 
(see e.g.~\cite{BFHSS08,CSL07,HH07,XZPCR09,ZPR09}).  Here, the given set of partial
orders consists of two subsets; one subset contains linearly ordered
votes and the other one completely unordered votes. Clearly, all
NP-hardness results would carry over from \textsc{Manipulation} to \textsc{Possible Winner}. However, whereas  the case of \emph{weighted voters} is settled by 
a full dichotomy for \textsc{Manipulation} for scoring
rules~\cite{HH07}, so far, for \emph{unweighted voters}  we are  only aware of one NP-hardness result  for a specially constructed scoring rule~\cite{XCP09}.
Indeed, the NP-hardness of \textsc{Manipulation} for Borda is a
prominent open question~\cite{XCP09,XZPCR09}. There are NP-hardness
results for \textsc{Manipulation} in the unweighted voter
case for several common voting rules
which are not  scoring rules~\cite{FHHR09tark,FHS08,XZPCR09}.  Another closely related
problem is \textsc{Preference Elicitation} (see
e.g.~\cite{CS02,CS05}). Here, the idea is to avoid that each voter has
to report his whole preference list, but to ask only for some part of
 the information that  suffices to determine a winner.

Now, let us briefly summarize the known results for \textsc{Possible
Winner} for scoring rules.  Correcting Konczak and Lang~\cite{KL05}
who claimed polynomial-time solvability for all scoring rules, Xia and
Conitzer~\cite{XC08} provided NP-completeness results for a class of
scoring rules, more specifically, for all scoring rules that have four
``equally decreasing score values'' followed by another ``strictly
decreasing score value''; we will provide a more detailed discussion
later. Betzler et al.~\cite{BHN09} studied the multivariate 
complexity of \textsc{Possible Winner} for scoring rules and other
types of voting systems, providing an NP-hardness proof
for~$k$-approval in case of only two partial votes. However, this 
NP-hardness result holds only if~$k$ is part of the input and does not
carry over for fixed values of~$k$. Furthermore, whereas the corresponding many-one reduction
relies on two partial votes, the construction used in this work makes
use of an unbounded number of partial votes and thus is completely
different.

Until now, the computational complexity of \textsc{Possible Winner}
was still open for a large number of naturally appearing scoring
rules. One such open case has been~$k$-approval for small values of $k$ which is motivated as
follows. A common way of voting for a board consisting of a small number, for example, of 
 five
members, is that every voter  awards one point each to five of the candidates
($5$-approval).  A second example is given by voting systems in which
each voter is allowed to specify a (small) group of favorites and a
(small) group of most disliked candidates. As final example, we
mention scoring rules that have decreasing differences between
successive score values as, for example, the scoring vector~$(2^m,
2^{m-1}, \dots, 0)$.

This work aims at a computational complexity dichotomy for {\it pure}
scoring rules. The class of pure scoring rules covers all of the
common scoring rules. It only constitutes some restrictions in the
sense that for different numbers of candidates the corresponding
scoring vectors cannot be chosen completely independently (see
 Section~\ref{sec:Prelim}). Our results can also be extended to broad classes of
``non-pure'' scoring rules, see Section~\ref{sec:hybrid}. Altogether,
we settle the computational complexity of {\sc Possible Winner} for
all pure scoring rules except the scoring rule defined
by~$(2,1,\dots,1,0)$. For plurality and veto, we provide
polynomial-time algorithms whereas for the remaining cases we show
NP-completeness.  Surprisingly, this includes the NP-hardness of
\textsc{Possible Winner} even for~$2$-approval. Our
NP-hardness result for 2-approval  has also been used to settle the
complexity of the
\textsc{Swap Bribery} problem~\cite{EFS09}.

\section{Preliminaries}
\label{sec:Prelim}

Let~$C = \{c_1, \dots, c_m\}$ be the set of {\it candidates}. A {\it vote} is a linear order
(i.e., a transitive, antisymmetric, and total relation) on~$C$. An $n$-voter
profile~$P$ on~$C$ consists of~$n$ votes~$(v_1, \dots, v_n)$ on~$C$. 
A \emph{voting rule}~$r$ is a function from the set of all profiles
on~$C$ to the power set of~$C$, that is~$r(P)$ denotes the set of winners.
{\it (Positional) scoring rules} are a special kind of voting rules. They are defined by scoring
vectors~$\overrightarrow{\al} = (\al_1,
\al_2, \dots, \al_m)$ with integers~$\al_1 \geq \al_2 \geq \dots \geq \al_m$, the {\it score values}. More specifically, we define that a scoring rule~$r$ consists of a sequence of scoring vectors $s_1, s_2, \dots$ such that for any~$i \in \mathbbm N_{>0}$ there
is a scoring vector~$s_i$ for $i$ candidates which can be computed in
time polynomial in~$i$.\footnote{For scoring rules that are defined
for a constant number of candidates, the \textsc{Possible Winner}
problem can be decided in polynomial time, see~\cite{CSL07,Wal07}.}
Here, we focus our attention on \emph{pure} scoring rules, that is
for every~$i\geq 2$, the scoring vector for~$i$ candidates
can be obtained from the scoring vector for~$i-1$ candidates by
inserting an additional score value at an arbitrary position
(respecting the described monotonicity). This definition includes all
of the common protocols like Borda or $k$-approval. We further assume
that~$\al_m =0$ and that there is no integer greater than one  that divides all score
values. This does not constitute a restriction since for every other 
voting system there must be an equivalent one that fulfills these
constraints~\cite[Observation~2.2]{HH07}. Moreover, we only consider
\emph{non-trivial} scoring rules, that is, scoring rules with~$\al_1
\neq 0$ for scoring vectors of every size.

For a vote~$v \in P$ and a candidate~$c \in C$, let the {\it
score}~$s(v,c)$ be defined by~$s(v,c) := \al_j$ where~$j$ is the
position of~$c$ in~$v$. For any profile~$P=\{v_1, \dots, v_n\}$,
let~$s(P,c) :=
\sum_{i=1}^{n}s(v_i,c)$. Whenever it is clear from the context which~$P$ we
refer to, we will just write~$s(c)$.  A scoring rule 
selects all candidates~$c$ as winners with maximum $s(P,c)$ over all candidates.

A {\it partial vote} on~$C$ is a transitive and antisymmetric relation
on~$C$. We use $>$ to denote the relation given between candidates in
a linear order and $\succ$ to denote the relation given between
candidates in a partial vote. Sometimes, we specify a whole subset of
candidates in a partial vote, e.g., $e \succ D$ for a candidate~$e \in C$ and a subset of candidates~$D\subseteq C$. Unless stated
otherwise, this notation means that $e \succ d$ for all~$d \in D$ and
there is no specified order among the candidates in~$D$. In contrast,
writing~$e > D$ in a linear order means that $e>d_1> \dots >d_l$ for
an arbitrary but fixed order of~$D=\{d_1, \dots, d_l\}$. A linear
order~$v'$ \emph{extends} a partial vote~$v$ if~$v
\subseteq v'$, that is, for any~$i,j \leq m$, from $c_i \succ c_j$ in~$v$ it follows that $c_i > c_j$ in~$v'$. 
Given a profile of partial votes~$P = (v_1, \dots, v_n)$ on $C$,
a candidate~$c \in C$ is a {\it possible winner} if there
exists an extension~$P'=(v'_1,
\dots, v'_n)$ such that each~$v'_i$ extends~$v_i$ and $c \in r(P') $.
The corresponding decision problem is defined as follows.

\begin{quote}
\textsc{Possible Winner}\\
\textbf{Given:} A set of candidates~$C$, a profile of partial votes~$P = (v_1, \dots, v_n)$ on $C$, and a distinguished 
candidate~$c \in C$.\\
\textbf{Question:} Is there an extension profile~$P'=(v'_1,
\dots, v'_n)$ such that each~$v'_i$ extends~$v_i$ and $c \in r(P')$ ?
\end{quote}

\noindent
This definition allows  that multiple candidates obtain the maximal score and we end up with a whole set of winners. If the possible winner~$c$ has to be  unique, one speaks of a possible {\it unique winner}, and the corresponding decision problem is defined analogously. All our results hold for both cases.

Several of our NP-hardness proofs rely on
reductions from the NP-complete \textsc{Exact Cover By 3-Sets} (X3C)
problem~\cite{GJ79}  defined as follows. Given a set of elements~$E=\{e_1, \dots, e_q\}$, a family of
subsets~$\mathcal{S} =\{S_1, \dots, S_t\}$ with~$|S_i|=3$ and $S_i
\subseteq E$ for~$1 \leq i \leq t$, it asks whether there is a subset~$\mathcal{S'} \subseteq
\mathcal{S}$ such that for every element~$e_j \in E$ there is exactly
one $S_i \in \mathcal{S'}$ with $e_j \in S_i$. In our NP-hardness proofs we need to describe the consequence of extending   partial votes for specific candidates. To this end, we say that a candidate~$c_i$ is \emph{shifted to the left (right)} by another candidate~$c_j$ when adding the constraint $c_i \succ c_j$ ($c_j \succ c_i$) to a partial vote.

In some of our theorems, we will need functions that map each instance of a certain problem~$\mathcal{P}$ to some natural number and  in some sense behave like a polynomial. For this sake, we call $$f: \{I \mid I \mbox{ is an instance of } \mathcal{P}\} \rightarrow \mathbbm{N}$$ a {\it poly-type function for~$\mathcal{P}$} if the function value~$f(I)$ is bounded by a polynomial in~$|I|$ for every input instance~$I$ of~$\mathcal{P}$. 

\section{General strategy}\label{sec:strategy}
This work aims at providing a dichotomy for \textsc{Possible Winner}
for practically relevant scoring rules. To this end, we will show the
following.

\begin{table}

\begin{center}
    \begin{tabular}{   l  l l  l}
   \toprule
    Scoring rule  & Result  &  \\
\midrule
Plurality and Veto & in P  & Proposition~\ref{prop:poly}, Section~\ref{sec:poly} \\
    different-type & NP-c (X3C) &  Theorem~\ref{theo:aaai}, Section~\ref{sec:different} 
    \\ 
   equal-type & NP-c (MC/X3C) & Theorem~\ref{theo:approval2}, Lemmata~\ref{lem:I} -- \ref{lem:IV}, Section~\ref{sec:2equal}  \\ 

  $\al_1> \al_2 =\al_{m-1}> 0$ & NP-c (X3C) &  Theorem~\ref{theo:3score}, Section~\ref{sec:3equal} \\
 \hspace{.5cm} and $\al_1 \neq  2 \cdot \al_2$\\
$(2,1,\dots,1,0)$&?&\\
    \bottomrule
    \end{tabular}
\end{center}
\caption{
Overview of results and outline of the work. Basically, we partition
the scoring rules into five different types according to the types of
algorithms or many-one reductions that are used to achieve the results. By
``different-type'' we denote all scoring vectors with an
unbounded number of different score values. By ``equal-type'' we denote all scoring vectors with an unbounded number of equal score values if not listed explicitly in another type. Reductions are from
\textsc{Exact Cover By 3-Sets} (X3C) or \textsc{Multicolored Clique}
(MC).}\label{tab:overview}
\end{table}

\medskip\noindent
\textbf{Theorem.} \textit{\textsc{Possible Winner} is NP-complete for all non-trivial pure scoring rules 
except plurality, veto, and scoring rules for which there is a constant~$z$ such that the produced  scoring
vector is~$(2,1,\dots,1,0)$ for every number of candidates greater than~$z$. For 
plurality and veto, \textsc{Possible Winner} is solvable in polynomial time.}
\medskip

The proof consists of several parts, see Table~\ref{tab:overview} for
an overview. The polynomial time results for plurality and veto are
based on flow computations. Regarding the NP-hardness results, we give
many-one reductions that work for  scoring rules that produce 
specific ``types of scoring vectors'' for an appropriate number of
candidates. We combine the single results  to obtain the
main result in Section~\ref{sec:main}. To this end, we have to take
into account that, in general, a scoring rule might produce different
types of scoring vectors for different numbers of candidates.

The basic observation to classify the scoring vectors is that a scoring
vector of unbounded size must have an unbounded number of different
 score values or an unbounded number of equal score values.  This leads
to the following strategy.  First, we show NP-hardness for all scoring
vectors having an unbounded number of different score values. To this
end, we generalize a many-one reduction due to Xia and
Conitzer~\cite{XC08}. Second, we deal with scoring vectors having an
unbounded number of equal score values. Here, we consider two
subcases, i.e., scoring vectors of type $\al_1> \al_2
=\al_{m-1}> 0$ but~$\al_1 \not= 2 \cdot \al_2$, and all remaining scoring vectors with an unbounded
number of equal score values.

Before stating the specific results, we give a construction scheme
that is used in all many-one reductions in this work.

\subsection{A General Scheme to Construct Linear Votes}
\label{sec:scheme}

In all many-one reductions presented in this work, one constructs a
partial profile~$P$ consisting of a set of linear orders~$V^l$ and a 
set of partial votes~$V^p$. The position of the
distinguished candidate~$c$ is already determined in every vote 
from~$V^p$, that is,~$s(P',c)$ is the same in every extension~$P'$ and thus is fixed.  The ``interesting'' part of
the reductions is given by the partial votes of~$V^p$ in combination with
upper bounds for the scores which the non-distinguished candidates can make in~$V^p$.  For
every candidate~$c' \in C \backslash
\{c\}$,  the \emph{maximum partial score}~$s^{\max}_p(c')$ is the
maximum number of points~$c'$ may make in~$V^p$ without beating~$c$
in~$P$. More precisely, for the unique winner case, $s_p^{\max}(c') =
s(P',c) - s(V^l,c')-1$ and, for the winner case,~$s_p^{\max}(c') =
s(P',c) - s(V^l,c')$ for any extension~$P'$ of~$P$. Since the maximum
partial scores can be adjusted to the unique and to the winner case,
all results hold for both cases.

In the following, we show that for all our reductions, there is an easy way to cast the linear votes
such that the maximum partial scores that are required in the reductions are
\emph{realized}. For every many-one reduction of this work, it will be easy
to verify that the underlying partial profile fulfills the following
two properties.\footnote{The only exception appears in the proof of
Theorem~\ref{theo:3score} and will be discussed there.}

\begin{quote}
 \textbf{Property~1}
 There is a ``dummy'' candidate~$d$ which cannot beat the distinguished candidate in any extension, that is,~$s^{\max}_p(d)
\geq \al_1 \cdot |V^p|$.
\end{quote}
 
\begin{quote}
\textbf{Property~2} For every~$ c' \in C \backslash\{c\}$, the maximum partial score~$s^{\max}_p(c')$ can be written as a sum of at most~$|V^p|$  integers from $\{\al_1, \dots, \al_{m}\}$. Formally, the definition of $s^{\max}_p(c')$ will be of the form $s^{\max}_p(c') = \sum_{j=1}^m n_j \al_j$ where $n_j \in \mathbbm{N}_0$ denotes how often the score value~$\al_j$ is added. We will always have that $\sum_{j=1}^{m} n_j \leq |V^p|$, that is, the total number of summands is at most the number of partial votes.
\end{quote}

The sets of linear votes which are necessary for the
reductions given in this paper can be obtained according to the
following lemma.

\begin{lemma}\label{lem:linearvotes}
Given a scoring rule~$r$, a set of candidates~$C$ with distinguished
candidate~$c \in C$, a set of partial votes~$V^p$ in which~$c$ is fixed,
and~$s^{\max}_p(c')$ for all~$ c' \in C \backslash\{c\}$, a  set of linear
votes that realizes the maximum partial scores for all candidates
can be constructed in time polynomial in~$|V^p|$ and~$m$  if Properties~1 and~2 hold.
\end{lemma}

\begin{figure}[t]
\center
$\begin{array}{lllllllllll}
v_1: & \quad \quad & c_1 & > &c_2 & > & \dots & > & c_{m-1} & > & c_m\\
v_2: & \quad \quad & c_2 & > & c_3 & > & \dots &> & c_{m} & > & c_1\\
\vdots &  \quad \quad & \vdots & & \vdots & & \vdots & & \vdots &  & \vdots \\
v_{m-1}: & \quad \quad & c_{m-1} & >& c_m & > & \dots &> & c_{m-3} & > & c_{m-2}\\
v_{m}: & \quad  \quad& c_m & >& c_1 & > & \dots & > & c_{m-2} & > & c_{m-1}\\
\end{array}$
\caption{Circular block for $c_1, c_2 \dots, c_m$ }\label{fig:perm}
\end{figure}

\begin{proof}
We are interested in ``setting'' relative score difference between the
distinguished candidate~$c$ and every other candidate. By inserting
one linear order we change the relative score difference between~$c$ and all
other candidates. To be able to change the relative score difference
only for $c$ and one specific candidate while keeping the relative
score difference of~$c$ and all other candidates, we will build~$V^l$ by sets of 
 circular shifts instead of single votes.  More precisely,
for a set of candidates~$\{c_1, c_2,  \dots, c_m\}$ , 
a \emph{circular block} consists of $m$ linear orders as given in Figure~\ref{fig:perm}.
Clearly, all candidates have the same score within a circular
block.

We start with the construction for the winner case and then explain
how to adapt it for the unique winner case. For the winner case
($s_p^{\max}(c') = s(P',c) - s(V^l,c')$ for any extension~$P'$), for each
candidate~$c' \in C\backslash \{c,d\}$ where $d$ denotes a dummy as
specified in Property~1,  add the following votes to the set of linear
votes~$V^l$. For each $n_j \neq 0$ as specified in Property~2,
construct $n_j$ circular blocks over~$C$ such that in one of the
linear orders of every block,~$c'$ sits on position~$j$ and~$d$ sits
on position~$m$. Exchange the places of~$c'$ and~$d$ in this linear
order and add the modified circular block to~$V^l$. Then, for one
block, $c'$ has lost~$\al_j$ points and gained~$\al_m = 0$ points
relative to~$c$.  Thus, in total, one has the situation that~$c$
and~$c'$ have exactly the same score if~$c'$ makes~$s^{\max}_p(c')$
points in~$V^p$. This settles the winner case. 
For the unique-winner
case, we additionally decrease the score of~$c'$ by the minimum of
$\{\al_i -\al_j \mid \al_i > \al_j \text{ and } i,j \in \{1,2, \dots,
m\}\}$.  This can be achieved by adding a circular block such that
in one of the linear orders of the block,~$c'$ sits on
position~$\al_i$ and~$d$ sits on position~$\al_j$, and by exchanging
the places of~$c'$ and~$d$ in this linear order.  Then,~$c$ beats~$c'$
if $c'$ makes at most~$s^{\max}_p(c')$ points in~$V^p$ and $c'$
beats~$c$, otherwise.  

Altogether, due to Property~2, we add at most~$|V^p|$ summands for
each candidate. Hence, so far, the number of linear votes is
bounded by $m^2\cdot (|V^p| +1)$ and can be constructed in polynomial
time.  It remains to adjust the maximum partial score of~$d$. Until
now, we added at most $m \cdot ( |V^p| +1)$ circular blocks. Thus,~$d$
can make at most $\al_1 \cdot m \cdot |V^p|$ points more than~$c$. By
adding~$m(|V^p| +1) + |V^p|$ further circular blocks for candidates
from~$C\backslash\{d\}$ that are inserted in the first~$m-1$
positions, while~$d$ is put on the last position in these
votes,~$s^{\max}_p(d)$ can be realized in polynomial time.
\qed
\end{proof}

\section{Plurality and Veto}
\label{sec:poly}

Employing network flows turned out to be useful to design algorithms for several voting
problems (see e.g.~\cite{Fal08,FHHR09jair}). Here, by using some flow
computations very similar to~\cite[Theorem~6]{BHN09}, we show the
following.

\begin{proposition}\label{prop:poly} \textsc{Possible Winner} can be solved in polynomial time for plurality and veto.
\end{proposition}

\begin{proof}

\begin{figure}

\begin{tabular}{llc}
&&\multirow{5}{*}{\resizebox{.4\textwidth}{!}{\input{possflownad.pstex_t}}}\\
$v_1: a \succ c \succ d, b \succ c$\\
$v_2: c \succ a \succ b$&{$\Rightarrow c >a>b>d$}\hspace{.3cm}\\
$v_3: a\succ d \succ b$&{$\Rightarrow c >a>d>b$}\\
$v_4: a \succ b \succ c$&\\
$v_5: a \succ c, b \succ d$\\
\end{tabular}

\caption{\textsc{Possible Winner} for plurality: The left-hand side shows an example for an election and  the right-hand side the corresponding flow network. The votes~$v_2$ and~$v_3$ can be extended  such  that~$c$ takes the first position. The position of the remaining candidates in~$v_2$ and~$v_3$ is not relevant; one possibility how to extend these votes is shown in the picture.}\label{fig:flow}
\end{figure}
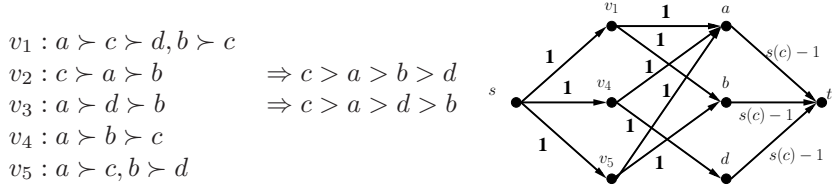

First, we give an algorithm for plurality.  Let~$P$ on~$C$ denote a
\textsc{Possible Winner}-instance with distinguished
candidate~$c$. Clearly, it is safe to set~$c$ to the first position in
all votes in which this is possible. Then the score of~$c$ is fixed at
the maximum possible value. We denote the partial votes of~$P$ in which the
first position is not taken by~$c$ as~$P_1$. Now, we can model the
problem as network flow as follows (see Figure~\ref{fig:flow}): The
flow network consists of a source node~$s$, a target node~$t$, one
node for every vote of~$P_1$, and one node for every candidate from~$C
\backslash\{c\}$. There are three layers of arcs:
\begin{enumerate}
\item an arc from~$s$ to every node corresponding to a vote in~$P_1$ with capacity one,
\item an arc from a node corresponding to~$v_j \in P_1$ to a node corresponding to a candidate~$c' \in C\backslash\{c\}$ with capacity one if and only if $c'$ can take the first position in an extension of~$v_j$, and
\item an arc from every node corresponding to~$c' \in C \backslash\{c\}$ to target~$t$ with capacity~$s(c)-1$.
\end{enumerate}

Now,~$c$ is a possible winner if and only if there is a flow of
size~$|P_1|$: The first layer simulates that the first position of
every partial vote from~$P_1$ has to be taken, the second layer that it
can only be taken by appropriate candidates, and the last one that the
score of every candidate will be lower than the score of~$c$.
Clearly, the flow network can be constructed in time polynomial
in~$|P_1|$ and an integral flow computation can be done in polynomial
time~\cite{CLRS01}.

For veto, we first fix~$c$ at the best (leftmost) possible position in every vote. This fixes the maximum score of~$c$. Then for every candidate~$c' \in C\setminus \{c\}$, let $z(c')$ denote
the minimum number of last positions that~$c'$ must take such that it
does not beat~$c$. Let $P_1$ denote the set of partial votes in which $c$ does not take the last position. 
 Again, we  model the
problem by  a flow network with source node~$s$,  target node~$t$, one node for every candidate from~$C
\backslash\{c\}$, and one node for every vote of~$P_1$. The arcs are as follows:
\begin{enumerate}
\item an arc from~$s$ to every node corresponding to~$c' \in C
\backslash\{c\}$ with capacity $z(c')$,
\item an arc from a node corresponding to ~$c' \in C
\backslash\{c\}$ to a node corresponding to~$v_j \in P_1$ with capacity one if and only if~$c'$ can take the last position in an extension of~$v_j$, and
\item an arc from every node corresponding to~$v_j \in P_1$ to target~$t$ with capacity~$1$.
\end{enumerate}
By similar arguments as for plurality, it follows that  $c$ is a possible winner if and only if there is a flow of size $\sum_{c' \in C \backslash\{c\}}z(c')$. 
\qed
\end{proof}

\section{An unbounded number of positions with different score values}\label{sec:different}

 Xia and Conitzer~\cite{XC08} developed a many-one reduction from \textsc{Exact
Cover By 3-Sets} showing that {\sc Possible Winner} is NP-complete for any
 scoring rule with scoring vectors which contain four consecutive,
 ``equally decreasing'' score values, followed by another strictly
 decreasing score value.  Using some additional gadgetry, we
 extend their proof to work for scoring vectors with an unbounded
 number of different, not necessarily equally decreasing score values.

We start by describing the basic idea employed in~\cite{XC08} (using a
slightly modified construction). Given an
X3C-instance~$(E,\mathcal{S})$, construct a partial profile~$P := V^l
\cup V^p$ on a set of candidates~$C$ where~$V^l$ denotes a set of
linear orders and~$V^p$ a set of  partial votes. To describe the basic
idea,  assume that there is a scoring vector with~$\al_1 > \al_2$ and 
 and the differences between the four following
score values are \emph{equally decreasing}, that is,
$\al_2-\al_3=\al_3-\al_4 =\al_4-\al_5$.  
Then, $C:= \{c,x,w\} \cup E$ where $E$ is the universe from the X3C-instance. 
The distinguished candidate is~$c$. The candidates whose
element counterparts belong to the set~$S_i$ are denoted by $e_{i1},
e_{i2}, e_{i3}$. The partial votes~$V^p$ consist of one partial vote~$v_i^p$ for every~$S_i \in \mathcal{S}$ which is given by
$$  x \succ e_{i1} \succ e_{i2} \succ e_{i3} \succ  C' \text{, }  w \succ  C'$$
with~ $C':= C\backslash \{x,e_{i1},e_{i2},e_{i3},w\} $.
Note that in~$v_i^p$, the positions of 
all candidates except $w,x,e_{i1},e_{i2},e_{i3}$ are fixed. More precisely,~$w$ has to be inserted between positions~$1$ and~$5$ maintaining the partial order $x \succ e_{i1} \succ e_{i2} \succ e_{i3}$.
%
%
 By setting the linear votes, the maximum partial scores are realized such that the following three
 conditions hold. 
\begin{itemize}
 \item For every \emph{element candidate} $e \in E$ one has the
 following. Inserting~$w$ behind~$e$ in two partial votes has the
 effect that $e$ would beat $c$, whereas when~$w$ is inserted
 behind~$e$ in at most one partial vote, $c$ still beats~$e$
 (Condition~1). Note that $e$ may occur in several votes at different
 positions, e.g.~$e$ might be identical with~$e_{i1}$ and~$e_{j3}$
 for~$i\neq j$. However, due to the condition of ``equally
 decreasing'' scores, ``shifting''~$e$ increases its score by the same
 value in all of the votes.  

\item The maximum partial score of~$x$ is
 set such that if takes more than~$|V^p|- |E|/3$ times the first position, then it would beat~$c$. That is,~$w$ must be inserted before~$x$ at least $|V^p| -|E|/3$ times
 (Condition~2).  

\item We set~$s_p^{\max}(w) = (|V^p|-|E|/3) \cdot \al_1 +
 |E|/3 \cdot \al_{5}$. This implies that if~$w$ is inserted
 before~$x$ in $|V^p|-|E|/3$ votes, then it must be inserted at the last
 possible position, that is, position~$5$, in all remaining votes
 (Condition~3).
\end{itemize}
Having an exact 3-cover for $(E,\mathcal{S})$, extend the partial votes as follows.

$
\begin{array}{llll}
v_i^p: x > e_{i1} > e_{i2} > e_{i3} > w > \dots & \text{ if }S_i \text{ is in the exact 3-cover}\\
v_i^p: w > x > e_{i1} > e_{i2} > e_{i3} >  \dots & \text{ if } S_i \text{ is not in the exact 3-cover.}\\
\end{array}
$

\noindent
Then, every element candidate~$e$ is shifted exactly once (in $v_i^p$ for $e \in S_i$, if $S_i$ is in the exact 3-cover) and thus is beaten by~$c$. It is easy to
verify that~$c$ beats~$w$ and~$x$ as well. In a yes-instance for
$(C,P,c)$, it follows directly from Condition~2 and~3 that~$w$ must
have the  position~$5$ in exactly~$|E|/3$ votes and the first 
position in all remaining partial votes. Since there
are $|E|/3$ partial votes such that three element candidates are shifted in 
each of them, due to Condition~1, every element candidate must appear in
exactly one of these votes. Hence,~$c$ is a possible winner in~$P$
if and only if there exists an exact 3-cover of~$E$.

By inserting further candidates, one can pad the construction
such that is also works if the equally decreasing score differences
appear at other positions~\cite{XC08}. Now, we consider the situation
in which no such equally decreasing score differences appear at
all. More precisely, we show how to extend the reduction to scoring
vectors with strictly, but not equally decreasing scoring values.  The
problem we encounter is the following: By sending candidate~$w$ to the
last possible position in the partial vote~$v_i^p$, each of the
candidates $e_{i1}, e_{i2}, e_{i3}$ improves by one position and
therefore improves its score by the difference given between the
corresponding positions. In \cite{XC08}, these differences all had the
same value, but now we have to deal with varying differences.  Since
the same candidate~$e \in E$ may appear in several votes at different
positions, e.g.~$e$ might be identical with~$e_{i1}$ and~$e_{j3}$
for~$i\neq j$, it is not clear how to set the maximum partial score
of~$e$.  Basically, to cope with this situation, we construct three partial
votes~$v_i^1,v_i^2,$ and~$v_i^3$ for every set~$S_i \in \mathcal{S}$
and permute the positions of the candidates $e_{i1}, e_{i2}, e_{i3}$
such that each of them takes a different position in $v_i^1$, $v_i^2$,
$v_i^3$. For example:

$v_i^1: \dots \succ x \succ e_{i1} \succ e_{i2} \succ e_{i3} \succ \dots$

$v_i^2: \dots \succ x \succ e_{i2} \succ e_{i3} \succ e_{i1} \succ \dots$

$v_i^3: \dots \succ x \succ e_{i3} \succ e_{i1} \succ e_{i2} \succ \dots.$

In this way, if the
candidate~$w$ is sent to the last possible position in all three
partial votes of a set~$S_i$, each of the candidates $e_{i1}, e_{i2},
e_{i3}$ improves its score by the same value. We only have to
guarantee that whenever~$w$ is sent back in the partial vote~$v_i^1$,
then it has to be sent back~$v_i^2$ and~$v_i^3$ as well. This is
realized by a gadget construction, which is the main technical
contribution of the following theorem.

\begin{theorem}An X3C-instance~$I$ can be reduced to a \textsc{Possible Winner}-instance for a scoring rule which produces a scoring vector having $f(I)$ positions with different score values. A suitable poly-type function~$f$  can be computed in polynomial time.\label{theo:aaai}
\end{theorem}

\begin{proof}

Given an X3C-instance~$(E,\mathcal{S})$ with~$\mathcal{S} = \{S_1,
\dots, S_t\}$ and~$S_i =\{e_{i1},e_{i2},e_{i3}\}$ for~$i\in \{1, \dots,
t\}$, construct a partial profile~$P$ on~$C$ as follows. The set of
candidates is defined as~$C:= \{x,w,c\} \uplus E \uplus D_{12} \uplus D_{13}
\uplus L$ (where $\uplus$ denotes the disjoint union), where~$E$ is the set of candidates that represent the elements of the universe of the X3C-instance, $D_{12} := \{d_1, \dots, d_t, h_1 , \dots, h_t\}$,
$D_{13}:=\{d_1', \dots, d_t', h_1' , \dots, h_t'\}$, and~$L := \{l_1,
\dots, l_t\}$. We define~$f\left((E,\mathcal{S})\right):= |C|$.  To ease the
presentation, we first assume that we have a strictly decreasing
scoring vector of size~$f\left((E,\mathcal{S})\right)$ and describe how to
generalize this at the end of the proof.
The partial profile
consists of a set of  partial votes~$V^p$ and a set of linear
votes~$V^l$. The partial votes are ~$V^p := \{v_i^1, v_i^2, v_i^3 \mid 1 \leq i \leq t\}$ with,

\smallskip
\noindent
for $1 \leq i \leq t-1$,\\
\arraycolsep0.05cm
$
\begin{array}{cccccccccccccccccccccc}
v_i^1: & x & e_{i1} & e_{i2} & e_{i3} & d_1 & \dots & d_i & h_{i+1} &\dots& h_t &  d_1' &\dots& d_i'&  h_{i+1}' &\dots& h_t' &\succ  C_i^1&, \hspace{.1cm} w \succ C_i^1\\ 
v_i^{2}: & x & e_{i2} & e_{i3} & e_{i1} & h_1 & \dots& h_i & d_{i+1} &\dots& d_t &  l_1 &\dots& &  \dots&  \dots &l_t &\succ  C_i^2&, \hspace{.1cm} w \succ C_i^2\\
v_i^{3}: & x & e_{i3} & e_{i1} & e_{i2} & l_1 & \dots  &  &\dots&  \dots & l_t &  h_1' &\dots& h_i'&  d_{i+1}' &\dots & d_t' &\succ  C_i^3&, \hspace{.1cm} w \succ C_i^3\\
\end{array}
$

\smallskip
\noindent
and\\
$
\begin{array}{cccccccccccccccccccccc}
v_t^1: & x & e_{t1} & e_{t2} & e_{t3} & d_1 &\dots& d_t &  d_1' &\dots& d_t' &\succ  C_t^1&, \hspace{.1cm} w \succ C_t^1\\ 
v_t^{2}: & x & e_{t2} & e_{t3} & e_{t1} & h_1 &\dots& h_t &  l_1 &\dots& l_t &\succ  C_t^2&, \hspace{.1cm} w \succ C_t^2\\
v_t^{3}: & x & e_{t3} & e_{t1} & e_{t2} & l_1 &   \dots & l_t &  h_1' &\dots& h_t' &\succ  C_t^3&, \hspace{.1cm} w \succ C_t^3\\
\end{array}
$

\begin{table}
\begin{tabular}{lll}
\toprule
&$s_p^{\max}(w)$ & $= (3t-q)\cdot \al_1 + q \cdot \al_{5+2t}$\\
& $s_p^{\max}(x)$ & $ = q \cdot \al_1 + (3t-q)\cdot \al_2$\\
$\forall e \in E$ &$s_p^{\max}(e)$ & $=(\al_2+\al_3+\al_4) + (n_e-1) \cdot (\al_3 +\al_4+\al_5) + \fixed(e)$\\
$\forall d_i$ & $s_p^{\max}(d_i)$ &$= q/3 \cdot \al_{4+i} + (t-q/3) \cdot \al_{5+i}+ \fixed(d_i)$\\
$\forall h_i$ & $s_p^{\max}(h_i)$ &$= q/3 \cdot \al_{4+i} + (t-q/3) \cdot \al_{5+i}+ \fixed(h_i)$\\
$\forall d_i'$ &  $s_p^{\max}(d_i')$ &$  = q/3 \cdot \al_{4+i+t} + (t-q/3) \cdot \al_{5+i+t} + \fixed(d_i')$\\
$\forall h_i'$ &  $s_p^{\max}(h_i')$ &$ = q/3 \cdot \al_{4+i+t} + (t-q/3) \cdot \al_{5+i+t} + \fixed(h_i')$\\
$\forall l_i$ &$s_p^{\max}(l_i)$ &$= 2t \cdot \al_1 + \fixed(l_i)$\\ 
\bottomrule
\end{tabular}
\caption{Maximum partial scores. Recall that $t =|\mathcal{S}|$, $q= |E|$, and $n_e =|\{S_i \in \mathcal S \mid e \in S_i\}|$.}
\label{tab:maxpar}
\end{table}

\smallskip
\noindent
where ``$\succ$'' signs are
 partially omitted and~$C_i^1,C_i^2$, and~$C_i^3$ denote the remaining candidates that are fixed in an arbitrary order, respectively. Now, we give some notation needed to define the maximum partial scores. For~$c' \in C \backslash\{c\}$, let $\fixed(c')$ denote the number of points which $c'$ makes in the partial votes in which the position of~$c'$ is already fixed. 
Let~$n_e$ denote the number of subsets with~$e \in S_i$  and $q =|E|$. Due to Lemma~\ref{lem:linearvotes}, one can set the  maximum partial scores as given in Table~\ref{tab:maxpar}. The particular partial scores will be explained within the proof of the following claim.

\smallskip\noindent
\textit{Claim}: Candidate~$c$ is a possible winner of~$P$ if and only if there is an exact 3-cover for~$(E,\mathcal{S})$.

\smallskip\noindent
``$\Leftarrow$'': Given an exact 3-cover~$\mathcal{S}'\subseteq
\mathcal{S}$, complete the votes in~$V^p$ in the following way: For
each~$S_i \in \mathcal{S}'$, place~$w$ in the last possible position
(i.e., position~$5+2t$) in the partial votes~$v_i^1, v_i^{2},$
and~$v_i^{3}$, and on the first position in the remaining partial
votes. Since~$|\mathcal{S'}| = q/3$, in the extension of the votes
from~$V^p$ ones has~$s(w) = (3t-q)\cdot \al_1 + q \cdot \al_{5+2t}
=s_p^{\max}(w)$ and $s(x) = q \cdot \al_1 + (3t-q)\cdot \al_2 =
s_p^{\max}(x)$. Furthermore, it is easy to see that~$s(l_i) < s_p^{\max}(l_i)$ for every~$i$. Every element candidate~$e$ is
shifted to the left in exactly three partial votes. More precisely, in
the three votes that correspond to~$S_i \in \mathcal S'$ with~$e \in
S_i$, it makes $\al_2 + \al_3 +\al_4$ points and $(n_e-1)
\cdot (\al_3 +\al_4+\al_5) + \fixed(e)$ points in the remaining
votes and thus does not beat~$c$.  Every candidate from~$D_{12}$ is
not ``fixed'' in exactly one vote of every triple corresponding to
an~$S_i$. More precisely, it can be shifted either in~$v_i^1$ or
in~$v_i^2$ and never in~$v_i^3$. Due to the insertion of~$w$, it is
shifted to position~$4+i$ in $q/3$ of the votes and takes
position~$5+i$ in the remaining $t-q/3$ non-fixed votes. Thus, it does
not beat~$c$. Analogously, every candidate from~$D_{13}$
makes~$\al_{4+i+t}$ points in~$q/3$ of the non-fixed votes and
$\al_{5+i+t}$ in the remaining~$t-q/3$ votes and hence does not
beat~$c$. Altogether,~$c$ beats every other candidate and wins.

\smallskip\noindent
``$\Rightarrow$'': Consider an extension of~$P$ in which~$c$ wins.
Due to its maximum partial score, candidate~$x$ can take the first
position only~$q$ times. Thus, it must be shifted $3t-q$ times to
position~$2$. Clearly, this is only possible if~$w$ is placed on the
first position in $3t-q$ votes. Then due to its maximum partial score,
$w$ can only be set to position~$5+2t$ in the remaining~$q$ votes. In
the following, we will show that for every~$i$,~$w$ takes
position~$5+2t$ in~$v_i^1$ if and only if it takes position~$5+2t$
in~$v_i^2$ if and only if it takes position~$5+2t$ in~$v_i^3$
(Observation~I). Then it follows that in the votes in which~$w$ takes
position~$5+2t$, the corresponding element candidates are shifted to
the left and obtain~$\al_2 +\al_3 +\al_4$ points each, whereas they
obtain~$\al_3 + \al_4 + \al_5$ points in the remaining corresponding
vote triples. Since each element candidate~$e_j$ can only obtain
$\al_2 +\al_3 +\al_4$ points exactly once (and the scoring values are
strictly decreasing), the set $\mathcal{S'} := \{ S_i\mid w \mbox{ is
placed on position $5+2t$ in } v_i^1\}$ must be an exact 3-cover
of~$E$.

It remains to settle Observation~I, which says that~$w$ behaves equally in
the votes corresponding to one subset. First, we argue that~$w$ must
be inserted at position~$5+2t$ in exactly $q/3$ votes of $V^p_1 :=
\{v_i^1 \mid 1\leq i\leq t\}$, $V^p_2 := \{v_i^2 \mid 1\leq i\leq
t\}$, and $V^p_3 := \{v_i^3 \mid 1\leq i\leq t\}$,
respectively. Assume that $w$ is inserted at position~$5+2t$ in more
than~$q/3$ votes of~$V^p_1$. Then,~$d_1$, which is not fixed in every
vote of~$V^p_1$, would beat~$c$. Analogously, if $w$ was inserted at
position~$5+2t$ in more than~$q/3$ votes of~$V^p_2$ or $V^p_3$,
 then $c$ would be beaten by~$h_1$ or $h_1'$, respectively. Now, we
have that $w$ must take position~$5+2t$ in $q$ votes and can take this
position in at most $q/3$ votes from~$V_i^p$, for every $i\in
\{1,2,3\}$ and thus must take this position in exactly $q/3$ votes of
~$V_1^p, V_2^p,$ and~$V_3^p$.

Second, we show that the candidates from~$D_{12}$ ensure that $w$
takes position~$5+2t$ in $v_i^1$ if and only if $w$ takes
position~$5+2t$ in $v_i^2$.  The proof is by contradiction. Assume
that there is an extension in which $w$ takes position~$5+2t$
in~$v_i^1$ and another position in~$v_i^{2}$ for any~$i$.  Since~$d_i$
and~$h_{i+1}$ have been shifted to the left in~$v_i^1$, each of them
can only be shifted to the left in at most~$q/3-1$ further votes. By
construction,~$v_i^{2}$ is the only vote of $V_1^p \cup V^p_2$ in
which neither~$d_i$ nor~$h_{i+1}$ is shifted to the left by
setting~$w$ to position~$5+2t$. However, since~$w$ can either take the
first or position~$2t+5$ in an extension (as argued above), it must
take the first position in~$v_i^{2}$. Now,~$w$ has to take the
position~$5+2t$ in~$2q/3-1$ further votes from~$V^p_1 \cup V^p_2$ and
thus in each of these votes~$w$ will either shift~$d_i$
or~$h_{i+1}$. Hence, either~$d_i$ or~$h_{i+1}$ must be shifted to the
left in more than~$q/3-1$ further votes and will beat~$c$, a
contradiction.  The other case ($w$ takes position~$5+2t$ in $v_i^2$
and another position in~$v_i^1$) follows in complete analogy by
considering~$h_i$ and~$d_{i+1}$.  One can show analogously
that the candidates of $D_{13}$ ensure that~$w$ takes position~$5+2t$
in~$v_i^1$ if and only if it takes the same position in~$v_i^3$. Thus,
Observation~I follows.

Now, one has that \textsc{Possible Winner} is NP-hard for all scoring
rules with a scoring vector of size~$f((E,\mathcal{S}))$ with strictly
decreasing score values. By using some simple padding, we extend the
result for the remaining cases, that is for scoring vectors of
size~$m'> f((E,\mathcal S))$ and$f((E,\mathcal{S}))$ different score
values. To this end, we introduce a set of $m'-f((E,\mathcal{S}))$
new dummy candidates and cast the linear votes such they cannot beat
the distinguished candidate in any extension.  The original candidates
from~$C$ are placed on positions endued with strictly decreasing
points, whereas the new candidates are placed on the remaining
positions. Then, if the positions of candidates get shifted (when~$w$
is inserted), the ``old'' candidates are affected in the same manner
as in the above construction and the theorem follows.
\qed
\end{proof}

\section{An unbounded number of positions with equal score values}\label{sec:equal}

In the previous section, we showed NP-hardness for scoring vectors
with an unbounded number of different score values.  In this section,
we discuss scoring vectors with an unbounded number of positions with
equal score value.  In the first subsection, we show NP-hardness for
\textsc{Possible Winner} for scoring vectors that fulfill~$\al_2 \neq
\al_{m-1}$, and, in the second subsection, we consider the special
case that~$\al_1 > \al_2 = \dots = \al_{m-1} > 0$. Note that these two cases cover all  scoring vectors with an unbounded number of equal score values (except plurality and veto): There are three ways to ``violate''~$\al_1 > \al_2 = \dots = \al_{m-1} > 0$. First, if one requires~$\al_1 = \al_2$, then one ends up with veto. Second, requiring~$\al_{m-1}=0$, one arrives at plurality. Third, requiring~$\al_2 \neq \al_{m-1}$, then one ends up with the other case that includes the famous examples 2-approval and $(m-2)$-approval.

\subsection{An unbounded number of  equal score values and~$\al_2 \neq \al_{m-1}$}\label{sec:2equal}

The scoring vectors considered in this subsection divide  into
two classes. First, there are at least two score values that are
greater than the ``equal score value''. Second, there are at least two
score values that are smaller than the ``equal score
value''. Formally, a size-$m$ scoring vector for the second class looks
as follows: there is an~$i$, with~$i<m-2$ and an ``unbounded''
integer~$x$ such that $\al_{i-x} =
\al_{i}>\al_{i+1}$. This property can be used to construct a basic
``logical'' tool used in the many-one reductions of this subsection:
For two candidates~$c,c'$, having $c \succ c'$ in a partial vote
implies that setting $c$ such that it makes less than~$\al_i$ points
implies that also $c'$ makes less than $\al_i$ points whereas all
candidates placed in the range between~$i-x$ and~$i$ make
exactly~$\al_i$ points. This can be used to model some implication of
the type ``$c \Rightarrow c'$'' in a vote. For $(m-2)$-approval, which
will play a prominent role for stating our results, this condition
means that~$c$ only has the possibility to make zero points in a vote
if also~$c'$ makes zero points in this vote whereas all other
candidates make one point.

Most of the reductions of this subsection are from the NP-complete
\textsc{Multicolored Clique (MC)} problem~\cite{FHRV09}:

\begin{quote}
\noindent
\textbf{Given:} An undirected graph~$G=(X_1 \cup X_2 \cup \dots \cup X_k, E)$ with  $X_i \cap X_j = \emptyset$  for $1 \leq i < j \leq k$ and the vertices of $X_i$ induce an independent set for $1 \leq i \leq k$.\\
\textbf{Question:} Is there a complete subgraph (clique) of size~$k$?
\end{quote}

\noindent
Here, $1,\dots ,k$ are considered as different colors. Then, the
problem is equivalent to ask for a \emph{multicolored clique}, that
is, a clique that contains one vertex for every color.  To ease the
presentation, for any~$1 \leq i \neq j \leq k$, we interpret the
vertices of $X_i$ as red vertices and write $r \in X_i$, and the
vertices of $X_j$ as green vertices and write $g \in X_j$.

Reductions from MC are often used to show parameterized hardness
results~\cite{FHRV09}. Intuitively, the different colors give some
useful structure to the instance.  The general idea is to construct
different types of gadgets. Here, the partial votes realize four kinds
of gadgets. First, gadgets that choose a vertex of every color (vertex
selection). Second, gadgets that choose an edge of every ordered pair
of colors, for example, one edge from green to red and one edge from
red to green (edge selection). Third, gadgets that check the
consistency of two selected ordered edges, e.g.\ does the chosen
red-green candidate refer to the same edge as the choice of the
green-red candidate (edge-edge match)? Finally, gadgets that check whether 
all edges starting from the same color start from the same vertex
(vertex-edge match).  Though reductions from MC have become a standard
tool to obtain hardness results, the reduction given here is not
straightforward. For example, we are not aware of any reduction in the
literature for which it is necessary to employ vertex-edge match
gadgets. 

We start by giving a reduction from MC that settles the NP-hardness of
\textsc{Possible Winner} for $(m-2)$-approval.  Then we describe how
the given construction can be generalized to work for most of the
cases considered in this subsection. The NP-hardness of the remaining
cases will be shown by reductions from \textsc{Exact Cover By 3-Sets}.

\begin{lemma}\textsc{Possible Winner} is NP-hard for $(m-2)$-approval.
\label{lem:approval}
\end{lemma}

\begin{proof}
Given an MC-instance~$G=(X, E)$ with $X=X_1 \cup X_2 \cup \dots \cup
X_k$. Let~$E(i,j)$ denote all edges from~$E$ between~$X_i$
and~$X_j$. Without loss of generality, we can assume that there are
integers~$s$~and~$t$ such that $|X_i|= s$ for $1\leq i \leq k$,
$|E(i,j)| = t$ for all $i,j$, and that $k$ is odd since every other
instance can be padded easily in this way.  We construct a partial
profile~$P$ on a set~$C$ of candidates such that the distinguished
candidate~$c \in C$ is a possible winner if and only if there is a
size-$k$ clique in~$G$.  The set of candidates~$C := \{c\} \uplus C_X
\uplus C_E \uplus D$, where $\uplus$ denotes the disjoint union, is
specified as follows:
\begin{itemize}
\item For $i \in \{1, \dots, k\}$, let $C_X^i :=\{r_1, \dots, r_{k-1} \mid r \in X_i\}$ and  $C_X := \bigcup_{i} C_X^i$.
\item  For $i,j \in \{1, \dots, k\}, i \neq j$, let $$C_{i,j} := \{ rg \mid \{r,g\} \in E(i,j), r \in X_i, \text{ and } g \in X_j\}$$ and $$C_{i,j}' := \{ rg' \mid \{r,g\} \in E(i,j), r \in X_i, \text{ and } g \in X_j\}.$$ Then, $C_E := (\bigcup_{i \neq j}C_{i,j}) \uplus (\bigcup_{i \neq j}C_{i,j}')$, i.e., for every edge~$\{r,g\}\in E(i,j)$, the set~$C_E$ contains the four candidates $rg, rg', gr, gr'$.
\item The set $D := D_X \uplus D_1 \uplus D_2$ is defined as follows. For~$i\in \{1,\dots,k\}$, $D_X^i := \{c_1^r,\dots, c_{k-2}^r \mid r \in X_i\}$ and $D_X := \bigcup_i D_X^i$.  For~$i \in \{1, \dots,k\}$,  one has $D_1^i := \{d_1^i, \dots, d_{k-2}^i\}$ and $D_1 := \bigcup_i D_1^i$. The set~$D_2$ is defined as $D_2 := \{d^i \mid i \in \{ 1,\dots, k\}\}$. 
\end{itemize}
We refer to the candidates of~$C_X$ as \emph{vertex-candidates}, to the candidates of~$C_E$ as \emph{edge-candidates}, and to the candidates of~$D$ as \emph{dummy-candidates}.

The partial profile~$P$ consists of a set of linear votes~$V^l$ and a
set of partial votes~$V^p$. In each extension of~$P$, the
distinguished candidate~$c$ gets one point in every vote from~$V^p$ (see
definition below). Thus, 
according to Lemma~\ref{lem:linearvotes},
 we can set the maximum partial scores as
follows. For every candidate~$d^i \in D_2$, $s_p^{\max}(d^i) = |V^p|-s+1$,
that is,~$d^i$ must get zero points (\emph{take a zero position}) in at least $s-1$ of the partial
votes. For every remaining candidate~$c' \in C \backslash(\{c\} \cup
D_2)$, $s_p^{\max}(c') = |V^p|-1$, that is,~$c'$ must get zero points
in at least one of the partial votes.

In the following, we define~$V^p := V_1 \cup V_2 \cup V_3 \cup
V_4$. For all our gadgets only the last positions of the votes are
relevant. Hence, in the partial votes it is sufficient to explicitly
specify the ``relevant candidates''.  More precisely, we define for
all partial votes that each candidate that does not appear explicitly
in the description of a partial vote is positioned before all
candidates that appear in this vote.

The partial votes of~$V_1$ realize the \textbf{edge selection
gadgets}. Basically, selecting an ordered edge~$(r,g)$ with~$\{r,g\}
\in E$ means to select the corresponding \emph{pair of
edge-candidates}~$rg$ and~$rg'$. The candidate~$rg$ is used for the
vertex-edge match check and~$rg'$ for the edge-edge match check. Now,
we give the definition of~$V_1$. For every ordered color pair~$(i,j),
i \neq j$, $V_1$ has $t-1$ copies of the partial vote $\{rg\succ rg'
\mid \{r,g\} \in E(i,j)\}$, that is, one  partial vote contains the
constraint~$rg\succ rg'$ for every~$\{r,g\} \in E(i,j)$.  The idea of
this gadget is as follows.  For every ordered color pair we have $t$
edges and $t-1$ corresponding votes. Within one vote, one pair of
edge-candidates can get the two available zero positions. Thus, it is
possible to set all but one, namely the selected pair of
edge-candidates, to zero positions.

The partial votes of~$V_2$ realize the \textbf{vertex selection
gadgets}. Here, we will use the $k-1$ candidates corresponding to a selected
vertex to do the vertex-edge match for all edges that are incident in
a multicolored clique. Formally, we set $V_2 := V_2^a \cup V_2^b$ as further defined in the following. Intuitively, in~$V_2^a$ we select a vertex and in~$V_2^b$, by a cascading effect,
we achieve that all $k-1$ candidates that correspond
to this vertex are selected.
  In $V_2^a$, for every color~$i$, we have $s-1$ copies of the partial
vote $\{ r_1\succ c_1^{r}\mid r \in X_i\}$.
In~$V_2^b$, for every color~$i$ and for every vertex~$r \in X_i$, we
have the following $k-2$~votes.

\begin{tabular}{llll}
For all \emph{odd} $z \in \{1,\dots, k-4\}$, &$v_z^{r,i}: \{c_z^r\succ c_{z+1}^r, r_{z+1}\succ r_{z+2}\}.$\\
 For all \emph{even} $z\in \{2, \dots,k-3\}$, & $v_z^{r,i}: \{c_z^r\succ c_{z+1}^r, d_{z-1}^i \succ d_z^i\}, $\\
 &$v_{k-2}^{r,i}: \{c_{k-2}^{r}\succ d_{k-2}^i, r_{k-1} \succ d^i\}.$
\end{tabular}

\smallskip
The partial votes of~$V_3$ realize the \textbf{vertex-edge match
gadgets}. For~$i,j \in \{1, \dots, k\}$, for~$j<i$,~$V_3$ contains the
vote $\{rg \succ r_j \mid \{r,g\} \in E, r \in X_i \text{, and } g \in X_j\}$ and,
for~$j>i$,~$V_3$ contains the vote $\{rg \succ r_{j-1} \mid \{r,g\} \in E, r \in X_i \text{, and } g \in X_j\}$.

The partial votes of~$V_4$ realize the \textbf{edge-edge match
gadgets}. For every unordered color pair~$\{i,j\}, i \neq j$ there is
the partial vote~$\{rg' \succ gr' \mid \{r,g\} \in E(i,j), r \in X_i, \text{ and } g \in X_j\}$.

This completes the description of the partial profile. Now, we  verify a property of the construction that is crucial to see the
correctness: In total, the number of zero positions available in the
partial votes is exactly equal to the sum of the minimum number of
zero position the candidates of~$C\backslash\{c\}$ must take such
that~$c$ is a winner.  We denote this property of the construction as
\emph{tightness}. 
To see the tightness property, we first compute the number of partial votes:

\begin{align}
|V_1|+|V_2|+|V_3|+|V_4|= \nonumber\\ 
  k(k-1)(t-1) + k(s-1) + ks(k-2)+ k(k-1)+ k(k-1)/2= \nonumber\\ 
 t(k^2-k) + s(k^2-k) + k^2/2 -3k/2.\label{numbervotes}
\end{align}

Regarding the number of zero positions that must be taken, we first compute the number of candidates for each subset:

\begin{itemize}
\item $|C_X| = sk(k-1)$,
\item $|C_E| = 2tk(k-1)$,
\item $|D_X| = sk(k-2)$, $|D_1|= k(k-2)$, and $|D_2|=k$.
\end{itemize}

The candidates of~$D_2$ must take at least $s-1$ zero positions and all other candidates at least one. Thus, in total the number of zero positions must be at least

\begin{align}
sk^2-sk+2tk^2-2tk+sk^2-2ks+k^2-2k+k(s-1)= \nonumber\\  2s(k^2-k)+2t(k^2-k)+k^2-3k.\label{number0}
\end{align}

Furthermore, there are two zero positions for every partial vote. It is
easy to verify that (\ref{numbervotes}) times two equals
(\ref{number0}). Hence, the tightness of the construction is shown.
 It directly follows that if there is a candidate that
takes more zero positions than desired, then~$c$ cannot win in this
extension since then at least one zero position must be ``missing''
for another candidate.

We can  now show the following claim to complete the proof.

\smallskip\noindent
\textit{Claim:} The graph~$G$ has a  clique of size~$k$ if and only if $c$ is a possible winner in~$P$.

\begin{figure}[t]

$\begin{array}{lllll}
V_1: && \cellcolor{yellow} \dots > rg > rg' &\cellcolor{yellow} \text{ for } i,j \in \{1,\dots, k\}, i \neq j, r \in X_i\backslash Q, \text{ and } g \in X_j\backslash Q \\
V_2^a: && \cellcolor{yellow} \dots > r_1 > c_1^{r} &\cellcolor{yellow} \text{ for } 1 \leq i \leq k \text{ and } r \in X_i\backslash Q\\
V_2^b: & v_z^{r,i}& \cellcolor{yellow}\dots > r_{z+1} >r_{z+2} &\cellcolor{yellow} \text{ for } 1 \leq i \leq k, r \in X_i\backslash Q \text{ for all } z \in \{1,3,5, \dots, k-4\}\\
       &v_z^{r,i}& \cellcolor{yellow} \dots > c_z^r >c_{z+1}^r &\cellcolor{yellow} \text{ for } 1 \leq i \leq k, r \in X_i\backslash Q \text{ for all } z \in \{2,4,6, \dots, k-3\}\\
   & v_{k-2}^{r,i}& \cellcolor{yellow} \dots > r_{k-1} > d^i &\cellcolor{yellow} \text{ for } 1 \leq i \leq k, r \in X_i\backslash Q\\
   &  v_z^{r,i}&\dots > c_{z}^r  > c_{z+1}^r & \text{ for } 1 \leq i \leq k, r \in X_i\cap Q \text{ for all } z \in \{1,3,5, \dots, k-4\}\\
       &v_z^{r,i}& \dots > d_{z-1}^i >d_{z}^i & \text{ for } 1 \leq i \leq k, r \in X_i \cap Q \text{ for all } z \in \{2,4,6, \dots, k-3\}\\
   & v_{k-2}^{r,i}& \dots > c_{k-2}^r > d^i_{k-2} & \text{ for } 1 \leq i \leq k, r \in X_i\cap Q\\
V_3: &&  \dots> rg>r_j & \text{ for } i,j \in \{1,\dots,k\}, j < i, r \in X_i \cap Q, \text{ and } g \in X_j \cap Q\\
&& \dots> rg>r_{j-1} & \text{ for } i,j \in \{1, \dots, k\}, j>i, r \in X_i \cap Q, \text{ and } g \in X_j \cap Q\\
V_4:&&  \dots > rg' > gr' &\text{ for } i,j \in \{1, \dots, k\}, i \neq j, r \in X_i\cap Q, g \in X_j \cap Q 
\end{array}
$
\caption{Extension of the partial votes for the MC-instance. Extensions in which candidates that do not correspond to the solution set~$Q$ take the zero positions are highlighted.}
\label{fig:MC}
\end{figure}

\smallskip\noindent
``$\Rightarrow$'' Given a multicolored clique~$Q$ of~$G$ of size~$k$. We refer to the vertices and edges belonging to~$Q$ as solution vertices and solution edges, respectively, and to the corresponding candidates as solution candidates.
Then, extend the partial profile~$P$ as given in Figure~\ref{fig:MC}.
In the following we argue that in the given extension every candidate
takes the required number of zero positions.

In~$V_1$, for every
 ordered color pair, all pairs of edge-candidates except the pair of solution edge-candidates
are set to the last two positions in one of the $t-1$ votes.

 In $V_2^a$ for every color~$i$, we set all candidates~$r_1$ that do not belong to the solution vertices and the corresponding~$c_1^r$ to  zero positions in one of the votes. 
 In
 $V_2^b$ for every non-solution vertex~$r \in X_i\backslash Q$
 we set the corresponding candidates~$r_{z+1}$ and~$r_{z+2}$ at zero positions
 in the votes~$v_z^{r,i}$ with odd index~$z\in\{1, \dots, k-4\}$.  In the votes with even index~$z \in \{2,\dots,k-3\}$,  we set the corresponding dummy candidates~$c_z^r, c_{z+1}^{r}$
 at zero positions. We further set the candidate~$r_{k-1}$ at a zero position in votes~$v_z^{r,i}$ for all the $s-1$ non-solution vertices of color~$i$, which 
 implies that the dummy candidate~$d^i$ is placed at $s-1$
 zero positions. Thus, we have ``enough'' zero positions for all the
 copies of the non-solution candidates, the corresponding dummy
 candidates~$\{c_1^r, \dots, c_{k-2}^r \mid r \in X_i \setminus Q\}$, and~$d^i$. The
 remaining votes of $V_2^b$ ``correspond'' to the gadgets for the
 solution vertices. Here, we set the candidate
 pairs~$c_z^r>c_{z+1}^r$ in the votes with odd index~$z\in \{1, \dots, k-4\}$ at position zero and
 the candidate pairs with candidates $d^i_p$ for $p=1, \dots, k-2$ to
 zero positions in the votes with even index. Thus, in~$V_2$, we have
 improved~$c$ upon all dummy candidates and upon all candidates
 corresponding to non-solution vertices, whereas each candidate
 corresponding to a solution vertex must still take a zero position.

Now, it remains to set every candidate corresponding to a solution vertex or a solution edge to a zero position in at least one vote. Due to
construction, for a solution edge~$\{r,g\}\in E$, the two
corresponding candidates~$rg'$ and~$gr'$ can be set to zero in the
corresponding vote of $V_4$. And, in $V_3$ the $k-1$ vertex-candidates belonging to every
solution vertex can be set to a zero position in combination with the
corresponding edge-candidate.  Thus, the distinguished candidate~$c$
is the winner of the described extension.

\smallskip\noindent
``$\Leftarrow$'' Given an extension of~$P$ in which~$c$ is a winner,
we show that the ``selected'' candidates must correspond to a
size-$k$~clique.  Recall that the number of zero positions that each
candidate must take is ``tight'' in the sense that if one candidate
gets an unnecessary zero position, then for another candidate there
are not enough zero positions left.

First (edge selection), for $i,j \in \{1, \dots, k\}, i \neq j$, we
consider the candidates of~$C_{i,j}$.  The candidates of $C_{i,j}$ can
take zero positions in one vote of $V_3$ and in $t-1$ votes
of~$V_1$. Since $|C_{i,j}| =t$ and in the considered votes at most one
candidate of~$C_{i,j}$ can take a zero position, every candidate of
$C_{i,j}$ must take one zero position in one of these votes. We refer
to a candidate that takes the zero position in $V_3$ as solution
candidate~$rg_{\sol}$. For every non-solution candidate~$rg \in C_{i,j}
\backslash \{rg_{\sol}\}$, its placement in~$V_1$ also
implies that~$rg'$ gets a zero position, whereas $rg_{\sol}'$ still
needs to take one zero position (which is only possible in~$V_4$).

Second, we consider the vertex selection gadgets. Here, analogously to
the edge selection, for every color~$i$, we can argue that in~$V_2^a$,
out of the set $\{r_1\mid r \in X_i\}$, we have to set all but one
candidate to a zero position. The corresponding \emph{solution vertex}
is denoted as~$r_{\sol}$. For every vertex~$r \in
X_i\backslash\{r_{\sol}\}$, this implies that the corresponding
dummy-candidate~$c_1^r$ also takes a zero position in~$V_2^a$. Now, we
show that in~$V_2^b$ we have to set all candidates that correspond to
non-solution vertices to a zero position whereas all candidates
corresponding to~$r_{sol}$ must appear only at one-positions. Since
for every vertex~$r \in X_i\backslash\{r_{\sol}\} $, the
vertex~$c_1^r$ has already a zero position in $V_2^a$, it cannot take
a zero position within $V_2^b$ anymore without violating the
tightness. In contrast, for the selected solution
candidate~$r_{\sol}$, the corresponding candidates~$c_1^{r_{\sol}}$
and~$r_{sol_1}$ still need to take one zero position. The only
possibility for~$c_1^{r_{\sol}}$ to take a zero position is within
vote~$v_1^{r_{\sol}, i}$ by setting $c_1^{r_{\sol}}$ and
$c_2^{r_{\sol}}$ to the last two positions. Thus, one cannot
set~$r_{\sol_2}$ and $r_{\sol_3}$ to a zero position
within~$V_2$. Hence, the only remaining possibility for ~$r_{\sol_2}$
and $r_{\sol_3}$ to get zero points remains within the corresponding
votes in~$V_3$. This implies for every non-solution vertex~$r$
that~$r_2$ and~$r_3$ cannot get zero points in~$V_3$ and thus we
have to choose to put them on zero positions in the vote~$v_1^{r,i}$ 
from~$V_2^b$. The same principle leads to a cascading effect in the
following votes of~$V_2^b$: One cannot choose to set
the candidates~$c_p^{r_{\sol}}$ for $p \in \{1, \dots, k-2\}$ to zero
positions in votes of $V_2^b$ with even index~$z$ and thus has to
improve upon them in the votes with odd index~$z$. This implies that all
vertex-candidates belonging to~$r_{\sol}$ only appear in one-positions
within $V_2^b$ and that all dummy candidates~$d^i_p$ for $p \in \{ 1,
\dots, k-2\}$ are set to one zero position. In contrast, for every
non-solution vertex~$r$, one has to set the candidates~$c_p^r$, $p \in \{2, \dots, k-2\}$, to zero positions in the votes with even index~$z$,
and thus in the votes with odd index~$z$, one has to set all
vertex-candidates belonging to~$r$ to zero positions. This further
implies that for every non-solution vertex in the last vote of~$V_2^b$
one has to set~$d^i$ to a zero position, and since there are exactly
$s-1$ non-solution vertices, $d^i$ takes the required number of zero
positions.
 Altogether, all vertex-candidates belonging to a solution vertex
 still need to be placed at a zero position in the remaining votes~$V_3
 \cup V_4$, whereas all dummy candidates of~$D$ and the candidates
 corresponding to the other vertices must have  taken enough zero
 positions.

Third, consider the vertex-edge match realized in~$V_3$. For $i,j \in
\{1, \dots, k\}, i \neq j$, there is only one remaining vote in which
$rg_{sol}$ with $r \in X_i$ and $g \in X_j$ can take a zero
position. Hence,~$rg_{sol}$ must take this zero-position. This implies
that the corresponding incident vertex-candidate~$x$ is also set to a
zero-position in this vote. If~$x \neq r_{sol_i}$, then~$x$ has
already a zero-position in~$V_2$. Hence, this would contradict the
tightness and $rg_{sol}$ and the corresponding vertex must
``match''. Furthermore, the construction ensures that each of the~$k-1$
candidates corresponding to one vertex appears exactly in one vote
of~$V_3$ (for each of the $k-1$ candidates, the vote corresponds to
edges from different colors). Hence,~$c$ can only be a possible winner
if a selected vertex matches with all selected incident edges.

Finally, we discuss the edge-edge match gadgets. In $V_4$, for~$i,j
\in \{1, \dots, k\}, i \neq j$, one still needs to set the solution candidates
from~$C_{i,j}$ to zero positions. We show that this can only be done
if the two ``opposite'' selected edge-candidates match each other. For
two such edges~$rg_{\sol}$ and $gr_{\sol}$, $r \in X_i, g
\in X_j$, there is only one vote in $V_4$ in which they can get a zero
position. If $rg_{\sol}$ and $gr_{\sol}$ refer to different edges,
then in this vote only one of them can get zero points, and thus the
other one still beats~$c$. Altogether, if~$c$ is a possible
winner, then the selected vertices and edges correspond to a
multicolored clique of size~$k$.
\qed
\end{proof}

By generalizing the  reduction used for Lemma~\ref{lem:approval}, one can show the following.

\begin{theorem}An MC-instance~$I$ can be reduced to a \textsc{Possible Winner}-instance for a scoring rule which produces a size-$m$ scoring vector that fulfills the following.  There is an~$i \leq m-1$ such that $\al_{i-x}= \dots = \al_{i-1} > \al_i$ with $x = f(I)$.  A suitable poly-type function~$f$  can be computed in polynomial time.\label{theo:approval2}
\end{theorem}

\begin{proof}
We describe how to modify the reduction given in the proof of
Lemma~\ref{lem:approval} to work for the considered cases. For this,
let~$P$ on~$C$ denote a partial profile as constructed in the proof of
Lemma~\ref{lem:approval}.  Since~$i\leq m-1$, the position~$i+1$ must
exist.  We set $x = f(I):= |C|-2$ and fill all positions smaller than~$i-x$ and
all positions greater than~$i+1$ with dummy candidates that are different from
candidates in~$C$ and that are beaten by~$c$ in every
extension. We distinguish the two subcases~$\al_i =
\al_{i+1}$~(1a) and $\al_i \neq \al_{i+1}$~(1b). 

 For the case~(1a), one can argue in complete analogy to
 Lemma~\ref{lem:approval} by ``identifying'' the two zero positions of
 Lemma~\ref{lem:approval} with position~$i$ and~$i+1$ and setting the
 maximum partial score as follows (which can be done without changing
 the partial votes due to Lemma~\ref{lem:linearvotes}). For all~$d^i
 \in D_2$, $s_p^{\max}(d^i) = (s-1) \cdot \al_i + (|V^p|-s +1) \cdot
 \al_{i-1}$ and for all~$c' \in C\backslash (\{c\} \cup D_2)$,
 $s_p^{\max}(c') = \al_{i} + (|V^p|-1) \cdot \al_{i-1}$.
 
 For~(1b), we
need to argue that the tightness argument still holds. For this, we
set the maximum partial scores as follows (which can be done without
changing the partial votes due to Lemma~\ref{lem:linearvotes}). For all~$d^i \in
D_2$, $s_p^{\max}(d^i) = (s-1)\cdot \al_{i+1} + (|V^p|-s+1) \cdot
\al_{i-1}$ and, for all~$c' \in C\backslash (\{c\} \cup D_2)$,
$s_p^{\max}(c') = \al_{i} + (|V^p|-1) \cdot \al_{i-1}$. Now, in any extension in which~$c$ wins, each
candidate in~$D_2$ must be placed at least $s-1$~times on position
$i+1$, and each of the other candidates must be placed on position~$i$
or~$i+1$ at least once. Then again, the number of positions~$i$
and~$i+1$ that still have to be assigned to candidates is exactly
equal to the number of candidates that need to take these positions,
hence, the tightness argument still holds. 
Thus, the correctness of the
modified reduction can be shown in complete analogy to
Lemma~\ref{lem:approval}.
\qed
\end{proof}

In the following, we consider scoring rules with an unbounded
number~$x$ of equal positions for which it holds that there is an~$i
\geq 2$ such that $\al_{i} > \al_{i+1} = \dots = \al_{i+x}$. Parts of
the results are based on further extensions of the MC-reduction used
to prove Lemma~\ref{lem:approval}. After that there still remain some
cases for which it seems even more complicated to adapt the
MC-reduction. However, for these cases we can make use of other
properties of the scoring rules and settle them by less involved
reductions from \textsc{Exact Cover by 3-Sets}. As we will see in Section~\ref{sec:main}, the
following Lemmata~\ref{lem:I}--\ref{lem:IV} cover all scoring vectors with~$i
\geq 2$ such that $\al_{i} > \al_{i+1} = \dots = \al_{i+x}$.

\begin{lemma}An MC-instance~$I$ can be reduced to a \textsc{Possible Winner}-instance for a scoring rule which produces a size-$m$ scoring vector that fulfills the following.  There is an~$i \geq 2$ such that $\al_{i} > \al_{i+1} = \dots = \al_{i+x}$  with $x = f(I)$ \textbf{and} there is a position~$j<i$ with $\al_{j} < 2 \al_{j+1}$.  A suitable poly-type function~$f$  can be computed in polynomial time.\label{lem:I}
\end{lemma}

\begin{proof}
We describe how to modify the MC-reduction given in the proof of
Lemma~\ref{lem:approval} to work for the considered case. For this,
let~$P$ on~$C$ denote a partial profile as constructed in the proof of
Lemma~\ref{lem:approval}.  First, we describe the construction
for~$j=i-1$, that is, one has~$\al_{i-1}<2 \al_i$.  We construct a
partial profile~$\widetilde{P}$ as follows. We set $x = f(I)= |C|-2$ and all
positions~$< i-1$ and~$> i+x$ are filled with dummy candidates that
are beaten by~$c$ in every extension.  The positions not filled with
dummies ``contain'' the partial votes of~$P$ in ``reverse'' order:
In~$P$ all relative orders are given for pairs of candidates. In~$\widetilde{P}$
we just ``flip'' every pair, for example, instead of having~$rg \succ
rg'$ we have~$rg' \succ rg$ in~$V_1$. We define that all
candidates that are not given explicitly are worse than the given
candidates in a vote (instead of being better). By flipping the order
of a pair, we adapt the ``logical implication'', for example, instead
of having ``if $rg$ makes zero points, then also $rg'$ makes zero
points'' in~$P$, we have ``if $rg$ makes~$\al_i$ points, then also
$rg'$ makes at least $\al_i$ points'' in~$\widetilde{P}$.  Furthermore, we set the
maximum partial scores to~$s_p^{\max}(d^i)= (s-1) \cdot \al_{i-1} +
(|V^p|-s+1)
\cdot \al_{i+1}$ for all~$d^i \in D_2$ and~$s_p^{\max}(c') = \al_{i-1} +
(|V^p|-1) \cdot \al_{i+1}$ for all~$c' \in C\backslash (\{c\} \cup
D_2)$. Note that since~$\al_{i-1}< 2 \al_i$, every candidate~$c'$ can
take either position~$i$ or position~$i-1$ in one of the partial
votes. Then, we can use a ``reverse'' tightness argument: Since the
positions~$i$ and~$i-1$ must be taken by two candidates in every
vote and every candidate can take at most one such position (or at
most $s-1$ such positions for candidates in~$D_2$, respectively), by
counting candidates and positions it holds that if every candidate
of~$D_2$ must make $\al_{i-1}$~points exactly $(s-1)$~times, then
every other candidate must make $\al_{i-1}$ or~$\al_i$~points exactly
once. Thus, it remains to show that
every~$d^i \in D$ must take position~$i-1$ in~$s-1$ of the
votes. Assume this is not the case, then there must be two
votes~$v_{k-2}^{r,i}$ and~$v_{k-2}^{r',i}$ with~$r\not = r'$ in which~$d^i$ does not
take position~$i-1$. Due to construction, the only remaining candidate
that can take this position in these votes is~$d^i_{k-2}$, but this is
not possible due to~$s^{\max}_p(d^i_{k-2})$. Hence, we can use a 
tightness argument analogously to Lemma~\ref{lem:approval}. Since we
also adapted the logical implication, the correctness follows in
complete analogy to Lemma~\ref{lem:approval}.

The remaining cases ($j<i-1$) follow by padding positions within the
gadgets. More precisely, replace each specified pair, e.g.\ $rg' \succ
rg$ by $rg' \succ rg \succ H$ with a dummy set~$H$ of size~$i-(j+1)$
and replace~$\al_{i-1}$ by $\al_j$ in the new definitions of the
maximum partial scores.
\qed
\end{proof}

So far, we settled the NP-hardness for scoring vectors with $i \geq 2$
such that $\al_{i} > \al_{i+1} = \dots = \al_{i+x}$ if there is a
position~$j<i$ with $\al_{j} < 2 \al_{j+1}$. Without the constraint
$\al_{j} < 2 \al_{j+1}$, it seems pretty complicated to adapt the
tightness property which is crucial for the MC-reduction. Fortunately,
the remaining cases have some different properties that allow to
settle them by less complicated reductions from \textsc{Exact
Cover By 3-Sets}. More precisely, in the following, we give three reductions
with increasing difficulty. (Although all three reductions are
self-contained, they might be easier to understand when reading them
in the given order.)

\begin{lemma}An X3C-instance~$I$ can be reduced to a \textsc{Possible Winner}-instance for a scoring rule which produces a size-$m$ scoring vector that fulfills the following.  There is an~$i \geq 2$ such that $\al_{i} > \al_{i+1} = \dots = \al_{i+x}$  with $x = f(I)$ \textbf{and} there is a position~$j<i$ with $\al_j \geq 3 \al_i$.  A suitable poly-type function~$f$ for X3C can be computed in polynomial time.\label{lem:II}
\end{lemma}

\begin{proof} Let~$(E,\mathcal{S})$ denote an
X3C-instance. Construct a partial profile~$P$ on a set of
candidates~$C$. The set~$C$ of candidates is defined by~$C := \{c\} \uplus S \uplus E \uplus H \uplus D$ where $c$ denotes the 
distinguished candidate~$c$,~$S:=\{s_z \mid S_z \in \mathcal{S}
\}$,~$E$ the set of candidates that represent the elements of the universe, and~$H$ and~$D$ contain disjoint candidates such that the following hold. We define~$H := \biguplus_{z=1}^{|\mathcal{S}|} H_z$ with~$|H_z|=i-j$ for all~$z \in \{1, \dots, |\mathcal{S}|\}$ needed to ``pad'' some positions relevant to the construction and $|D| = m- |S| -|E|-|H|-1 $ needed to pad irrelevant positions. We refer to the candidates from~$S$ as {\it subset candidates} and to the candidates from~$E$ as {\it element candidates}.
 Set $f((E,\mathcal{S})) := |C\setminus D|- (i-j)$.  For~$1 \leq z \leq
 |\mathcal S |$, let~$S_z = \{e_{z1},e_{z2},e_{z3}\}$. The partial
 profile~$P$ consists of a set of linear votes and a set of partial
 votes~$V^p$.  In all votes of~$V^p$, we pad all irrelevant positions,
 i.e.~all positions smaller than~$j$ and greater than~$j-1 +|C
 \setminus D|$ by fixing candidates from~$D$ (omitted in the further
 description).  The set~$V^p$ consists of~$|\mathcal S | -|E|/3$ copies
 of the vote $$s_1\succ H_1 \succ C\backslash(S \cup H), s_2\succ H_2
 \succ C\backslash(S \cup H), \dots, s_{|\mathcal{S}|} \succ
 H_{|\mathcal{S}|} \succ C\backslash(S \cup H) $$ denoted as~$V_1^p$
 and the following three votes, denoted as~$V_2^p(z)$, for every~$s_z \in
S$

$\begin{array}{lll}
v_z^1: & H_1 \succ \{s_z,e_{z1}\} \succ C\backslash(\{s_z, e_{z1}\} \cup H_1),\\
v_z^2: & H_1 \succ \{s_z,e_{z2}\} \succ  C\backslash(\{s_z, e_{z2}\} \cup H_1) \text{,and }\\
v_z^3: & H_1 \succ \{s_z,e_{z3}\} \succ  C\backslash(\{s_z, e_{z3}\} \cup H_1).\\
\end{array}$

\noindent
The basic idea of this construction is that in~$V_1^p$ one has to set all but $|E|/3$ ``subset'' candidates to position~$j$ whereas the remaining candidates will be able to take a position greater than~$i$ in all votes from~$V_1^p$.  Therefore, the remaining $|E|/3$ subset candidates can  make $\al_j-\al_{i+1}$  points more than the other candidates within the remaining votes. This  will enable them to shift their corresponding element candidates to position~$i+1$ by taking position~$i$. Since $\al_j > 3\cdot \al_i$, they will be able to shift all three element candidates, respectively. To realize the basic idea, we  adapt the maximum partial scores appropriately. For $e \in E$, let~$n_e$ denote the number of subsets in $\mathcal{S}$ which contain~$e$. Then according to Lemma~\ref{lem:linearvotes}, we can cast the linear votes such that the following holds:
\begin{itemize}
\item $s_p^{\max}(s_z) = \al_j + (|V^p|-1) \cdot \al_{i+1}$, for all~$s_z \in S$,
\item  $s_p^{\max}(e) = (n_e-1) \cdot \al_i + (|V^p|-n_e+1) \cdot \al_{i+1}$, for all~$e\in E$, and
\item all other candidates are beaten by~$c$ in every extension.
\end{itemize}

We show that~$c$ is a possible winner in~$P$ if
and only if there is an exact 3-cover for~$(E,\mathcal{S})$:

Assume there is an exact 3-cover~$Q$. Then one extends~$P$ by setting
each~$s_z$ with~$S_z \notin Q$ at position~$j$ in one vote
from~$V_1^p$ and the corresponding candidates from~$H_z$ to the
positions~$j+1, \dots, i$ in the same vote. Furthermore, set~$s_z$ to
position~$i+1$ in~$v_z^1,v_z^2$, and~$v_z^3$. Now, we have that
every~$s_z$ with $S_z \notin Q$ takes position~$j$ in one vote and a
position greater than~$i$ in all remaining votes and thus is beaten
by~$c$. This also means that in~$V_1^p$ all positions~$\leq i$ are filled and
thus every candidate~$s_z$ with~$S_z \in Q$ takes a position greater
than~$i$ in all votes from~$V_1^p$. Thus, the remaining votes can be
extended by setting every~$s_z$ with~$S_z \in Q$ to  position~$i$
in~$v_z^1,v_z^2$, and~$v_z^3$. Since~$\al_j \geq 3 \al_i$, the maximum
partial score of~$s_z$ is not exceeded. Because $Q$ is an exact
3-cover, all element candidates are shifted to  position~$i+1$ in one vote 
 and thus are beaten by~$c$. Hence,~$c$ is a winner in the described
extension.

For the other direction, consider an extension of~$P$ in which~$c$
wins. Due to construction, in~$V_1^p$ only subset candidates from~$S$
can take position~$j$. Because of the maximum partial scores,
position~$j$ must be taken by different candidates from~$S$ in
the~$|\mathcal{S}|-|E|/3$ votes of~$V^1_p$. We denote these candidates
as non-solution candidates and the remaining~$|E|/3$ candidates
 from~$S$ as solution candidates. Due
to~$s_p^{\max}(s_z)$, every non-solution candidate must take
position~$i+1$ in all remaining votes and thus the corresponding
element candidates must make~$\al_i$ points in the corresponding
votes. Hence, there remain only~$|E|/3$ solution candidates that have
to ``shift'' the~$|E|$ element candidates to position~$i+1$. Since every
solution candidate can shift at most~3 candidates, the solution
candidates must correspond to an exact 3-cover.
\qed
\end{proof}

In the following lemma, we consider a more specific type of scoring
vector in the sense that there are only two score values greater than
zero. This restriction allows us to find an easy way to ``lift'' the
condition ``$\al_j \geq 3 \cdot \al_i$'' for two special types of
scoring rules that will be sufficient for the proof of the main result
in Section~\ref{sec:main}. Compared to the reduction from the previous
lemma, for the following cases we also choose a set of ``solution
subset candidates'' within the first part of the partial votes, but we
will need some additional gadgetry to be able to ``shift'' the
corresponding element candidates.

\begin{lemma}An X3C-instance~$I$ can be reduced to a \textsc{Possible Winner}-instance for a scoring rule which produces a size-$m$ scoring vector ~$(\al_1,\al_2,0,\dots, 0)$ with $3\al_2 >\al_1>2 \al_2$ and  $m =f(I)+2$.
  A suitable poly-type function~$f$  can be computed in polynomial time.\label{lem:III}
\end{lemma}

\begin{proof}
Let~$(E,\mathcal{S})$ denote an
X3C-instance. Construct a partial profile~$P$ on a set of
candidates~$C$ as follows. The set of candidates consists of a
distinguished candidate~$c$, a set~$S:=\{s_i  \mid S_i \in \mathcal{S}
\}$ (the subset candidates), a set~$D :=\{d_i \mid S_i \in \mathcal{S}
\}$,  the set~$E$  (the element candidates), a candidate~$x$, and~$H := \{h_1, \dots, h_{|\mathcal{S}|} \}$. Set $f((E,\mathcal{S})) := |C|-2$. 
 For~$1 \leq i \leq |\mathcal S |$,
let~$S_i = \{e_{i1},e_{i2},e_{i3}\}$. The partial profile~$P$ consists
of a set of linear votes and a set of partial votes~$V^p$.
The set~$V^p$ consists of~$|\mathcal{S}|-|E|/3$ copies of the vote
$$s_1\succ h_1 \succ C\backslash (S \cup H ),s_2\succ h_2 \succ C\backslash (S \cup H ), \dots,  s_{|\mathcal{S}|}\succ h_{|\mathcal{S}|} \succ C\backslash (S \cup H )$$
denoted as~$V_1^p$ 
and the following three votes
for every~$S_i \in \mathcal{S}$

$\begin{array}{lll}
v_i^1:   & d_i \succ e_{i1} \succ C\backslash \{d_i,e_{i1},s_i\}, s_i \succ C\backslash \{d_i,e_{i1},s_i\}\\
v_i^2: & x \succ \{d_i,e_{i2}\} \succ C\backslash \{d_i,e_{i2},x\}\\
v_i^3: & x \succ \{d_i,e_{i3}\} \succ C\backslash \{d_i,e_{i3},x\}\\
\end{array}$

Let~$n_e$ denote the number of subsets in which~$e$ occurs. Then, due to Lemma~\ref{lem:linearvotes}, we can set the maximum partial scores as follows:
\begin{itemize}
\item $s_p^{\max}(s_i) = \al_1$ for all~$s_i \in S$,
\item $s_p^{\max}(d_i) = 3 \cdot \al_2$ for all~$d_i \in D$, 
\item  $s_p^{\max}(e) = (n_e-1) \cdot \al_2$ for all~$e \in E$, 
\item all other candidates are beaten by~$c$ in every extension.
\end{itemize}

We show  that~$c$ is a possible winner in~$P$ if
and only if there is an exact 3-cover for~$(E,\mathcal{S})$:

Assume there is an exact 3-cover~$Q$
for~$(E,\mathcal{S})$. Then we extend~$P$ as follows. For
every~$S_i\notin Q$, $s_i$ takes position~1 and~$h_i$ takes
position~$2$ in one vote from~$V_1^p$ and $s_i$ takes position~3 in
$v_i^1$. The corresponding~$d_i$ takes position~3 in~$v_i^2$
and~$v_i^3$. Clearly, for~$S_i \notin Q$, ~$s_p^{\max}(s_i)$ is not
exceeded, $s_p(d_i)=\al_1 <3\al_2 =s_p^{\max}(d_i)$, and
within~$V_1^p$ all first positions are fixed. For every solution
set~$S_i \in Q$, we set~$s_i$ to a position greater than~2  in all votes
from~$V_1^p$ and to the first position in~$v_i^1$. Since this implies
that~$d_i$ takes the second position in~$v_i^1$, this enables us to
set~$d_i$ to the second position in~$v_i^2$ and~$v_i^3$ without
violating~$s_p^{\max}(d_i)$. Since~$Q$ is an exact 3-cover, all
corresponding element candidates are shifted to the third position
once and for every element candidate the maximum partial score is not
exceeded. Hence,~$c$ is a winner.

To see the other direction, assume there is an extension in which~$c$
wins. In~$V_1^p$, the first positions can only be taken by candidates
from~$S$. Since each~$s_i \in S$ can get~$\al_1$ points exactly once,
$|\mathcal{S}|-|E|/3$ different subset candidates from~$S$ have to be
placed on the first position. Let the set consisting of these candidates be denoted by~$S'$. Every
candidate~$s_i$ from~$S'$ has exploited its maximum partial score and
therefore has to be placed on the third position in~$v_i^1$. This
implies that the corresponding candidate~$d_i$ takes the first
position in~$v_i^1$. Since~$\al_1 > 2\al_2$
and~$s_p^{\max}(d_i)=3\al_2$,~$d_i$ has to take the third position in
both~$v_i^2$ and~$v_i^3$. Hence, for~$s_i \in S'$, the corresponding
element candidates~$e_{i1}, e_{i2}, e_{i3}$ receive~$\al_2$ points
each. However, each of the element candidates from~$E$ has to be
placed on position~$3$ at least once due to its maximum partial
score. This can only be in the remaining partial votes, that is, all
$v_i^1,v_i^2,v_i^3$ with $s_i \in S \setminus S'$. Since $|S \setminus
S'| = |E|/3$, one must shift one element candidate in each of these
votes. For this, the only possibility is to set every~$s_i \in S
\setminus S'$ to position~1 in~$v_i^1$, and the corresponding
candidate~$d_i$ takes the second position in~$v_i^2$ and~$v_i^3$.  Since~$c$ wins, all~$|E|$ element candidates must get shifted to position~$3$.  Hence, $S\setminus S'$ corresponds to an exact 3-cover of~$(E,\mathcal{S})$.
\qed
\end{proof}

Finally, we settle the NP-hardness for a specific scoring vector.

\begin{lemma}An X3C-instance~$I$ can be reduced to a \textsc{Possible Winner}-instance for a scoring rule which produces a size-$m$ scoring vector ~$(2,1,0,\dots, 0)$ for $m =f(I)+2$.
  A suitable poly-type function~$f$  can be computed in polynomial time.\label{lem:IV}
\end{lemma}

\begin{proof}

Let~$(E,\mathcal{S})$ denote an
X3C-instance. Construct a partial profile~$P$ on a set of
candidates~$C$ as follows. The set of candidates consists of a
distinguished candidate~$c$, a set~$S:=\{s_i \mid S_i \in \mathcal{S}
\}$ (the subset candidates), $D:=\{d_i \mid S_i \in \mathcal{S}
\}$, $T:=\{t_i \mid S_i \in \mathcal{S}
\}$, $E$ (the element candidates), a candidate~$y$, and $X:= \{x_1, \dots, x_{|\mathcal{S}| - |E|/3}\}$. Set $f((E,\mathcal{S})) := |C|-2$. 
 For~$1 \leq i \leq |\mathcal S |$,
let~$S_i = \{e_{i1},e_{i2},e_{i3}\}$. The partial profile~$P$ consists
of a set of linear votes and a set of partial votes~$V^p$.
The set~$V^p := V^p_1 \cup V^p_2 \cup V^p_3$ is further defined as follows. 
The set~$V_1^p$ consists of $|\mathcal{S}|-|E|/3$ copies of the partial vote
$$s_1\succ t_1 \succ C\backslash (S\cup T),s_2\succ t_2 \succ C\backslash (S\cup T), \dots, s_{|\mathcal{S}|}\succ t_{|\mathcal{S}|} \succ C\backslash (S\cup T). $$
The set~$V_2^p$ consists of $|\mathcal{S}|-|E|/3$ copies of the partial vote
$$y\succ T \succ C\backslash (T \cup \{y\}) $$
and $V_3^p$ contains the following three votes
for every~$S_i \in \mathcal{S}$

$\begin{array}{lll}
v_i^1:   & d_i \succ e_{i1} \succ C\backslash \{d_i,e_{i1},s_i\}, s_i \succ C\backslash \{d_i,e_{i1},s_i\}\\
v_i^2: & y \succ \{d_i,e_{i2}\}\succ C\backslash \{d_i,e_{i2},y\}\\
v_i^3: &  \{t_i,e_{i3}\}\succ C\backslash (\{t_i,e_{i3}\} \cup X)\\
\end{array}$

Let~$n_e$ denote the number of subsets in which~$e$ occurs
and~$n_{e,3}$ the number of subsets in which $e$ is denoted
as~$e_{i3}$ for~$i \in \{1, \dots, |S|\}$. Then, using Lemma~\ref{lem:linearvotes}, we set the
maximum partial scores as follows:
\begin{itemize}
\item $s_p^{\max}(s_i) =s_p^{\max}(t_i) =s_p^{\max}(d_i) = 2$ for $i \in \{1, \dots,|\mathcal{S}|\} $
\item $s_p^{\max}(x_i) = 1$ for $i \in \{1, \dots, |\mathcal{S}|-|E|/3\}$
\item  $s_p^{\max}(e) = 2n_{e,3} +(n_e-n_{e,3})-1$ for $e \in E$
\item the candidate~$y$ is  beaten by~$c$ in every extension
\end{itemize}

We show  that~$c$ is a possible winner in~$P$ if
and only if there is an exact 3-cover for~$(E,\mathcal{S})$:

\begin{table}
$
\begin{array}{lllll}
V_1^p:& & \cellcolor{yellow}  s_i > t_i > \dots  &\cellcolor{yellow} \text{ for } S_i \notin Q \\
V_2^p: && \cellcolor{yellow}  y > t_i > \dots  &\cellcolor{yellow} \text{ for } S_i \notin Q \\
V_3^p:& v_i^1 &  \cellcolor{yellow}  d_i > e_{i1} > s_i > \dots  &\cellcolor{yellow} \text{ for } S_i \notin Q \\
& v_i^2 &  \cellcolor{yellow}  y > e_{i2}> d_i > \dots  &\cellcolor{yellow} \text{ for } S_i \notin Q \\
& v_i^3 &  \cellcolor{yellow}  e_{i3} > x_q > t_i \dots  &\cellcolor{yellow} \text{ for } S_i \notin Q \text{ and different } q \\
& v_i^1 &    s_i >  d_i > e_{i1}> \dots  & \text{ for } S_i \in Q \\
& v_i^2 &    y >  d_i > e_{i2}> \dots  & \text{ for } S_i \in Q \\
& v_i^3 &    t_i >   e_{i3}> \dots  & \text{ for } S_i \in Q \\
\end{array}
$
\caption{Extension for the X3C-reduction for the case $(2,1,0, \dots)$. The remark  ``different~$q$'' means that for $i \neq i'$ with~$S_i \notin Q$ and~$S_{i'} \notin Q$ one chooses two different candidates from~$X$. Extensions corresponding to non-solution candidates are highlighted.}
\label{tab:extX3C}
\end{table}

Assume there is an exact 3-cover~$Q$
for~$(E,\mathcal{S})$. Then we extend~$P$ as given in
Table~\ref{tab:extX3C}. For every~$S_i\notin Q$, $s_i$ takes the first
position in one vote from~$V_1^p$ and makes zero points in all
remaining votes. The corresponding~$t_i$ takes the second position in
one vote from~$V_1^p$ and one vote from~$V_2^p$ and makes zero points
in all remaining votes. Hence, $c$ beats these $s_i$ and $t_i$ and the
votes from $V_1^p$ and $V_2^p$ are fixed. For every~$S_i
\notin Q$, we extend~$v_i^3$ by setting a different candidate from~$X$
at the second position such that none of them is put on this position twice, and hence~$c$ also beats every candidate
from~$X$. For every~$S_i \in Q$,~$d_i$, $t_i$ and~$s_i$ make exactly 2 points
in~$V_3^p$ and thus are beaten by~$c$ as well. It remains to consider
the element candidates. To this end, note that a candidate~$e \in E$
is beaten by~$c$ if there is an~$i$ such that~$e$ takes position~3
in~$v_i^1$ or~$v_i^2$ or takes position~2 in~$v_i^3$. Since~$Q$ is
an exact 3-cover and all candidates corresponding to subsets from~$Q$
are shifted to the right in one vote,~$c$ wins in the given extension.

To see the other direction, assume there is an extension in which~$c$ wins. Let~$G^1 := \{v_i^1 \mid 1 \leq i \leq |\mathcal{S}|\}$, $G^2:= \{v_i^2\mid 1 \leq i \leq |\mathcal{S}|\} $, and $G^3:= \{v_i^3\mid 1 \leq i \leq |\mathcal{S}|\} $. We start by
arguing that at most $2/3 \cdot |E|$ candidates from~$E$ can make zero points in a vote from~$G^1 \cup G^2$.  For any~$i$, at most two element
candidates, namely~$e_{i1}$ and~$e_{i2}$ can make zero points in~$G^1
\cup G^2$. More precisely, due to~$s_p^{\max}(d_i)$, if~$s_i$ takes the first position in
$v_i^1$, then $e_{i1}$ and~$e_{i2}$ can take the third position and if
$s_i$ takes the second position, then only $e_{i1}$ can be shifted to
the third position, since~$d_i$ takes the first position in~$v_i^1$ and has exploited its maximum partial score. Thus, the number of points that all candidates
from~$S$ can make within~$V_3^p$ is an upper bound for the number of
element candidates that can be shifted. Since only candidates from~$S$
can take the first positions in~$V_1^p$, $|V_1^p|=
|\mathcal{S}|-|E|/3$, and~$s_p^{\max}(s_i)=2$, the candidates from~$S$
can make at most~$2/3 |E|$ points in~$V_3^p$. Thus, there are at most $2/3 |E|$
element candidates that can take a position with zero points in
$G^1\cup G^2$. Thus, due to~$s_p^{\max}(e)$, in~$G^3$ one must shift
(at least) $|E|/3$ candidates to the second position (\textbf{Observation~1}). In the
following, we show that the only way to do so leads to an extension in
which exactly $|E|/3$ candidates~$s_i$ from~$S$ make zero points
in~$V_1^p$ and the corresponding~$t_i$ make zero points in~$V_1^p \cup
V_2^p$ whereas all other candidates from~$S\cup T$ have already
accomplished their maximum partial score in~$V_1^p\cup V_2^p$ (\textbf{Claim~1}). This means
that the element candidates that are shifted to the right correspond to
exactly $|E|/3$ subsets~$S_i \in \mathcal{S}$. Since every element
candidate must be shifted at least once, these subsets must form an
exact 3-cover in~$(E,\mathcal{S})$.

We use a tightness criterion (analogously to the MC-reduction from
Lemma~\ref{lem:approval}) to prove Claim~1. To this end, we show that the score
of all positions that must be filled equals the sum of the maximum
partial scores of all candidates. Again, it directly follows that a
candidate~$c' \in C\backslash \{c\}$ cannot make less
than~$s_p^{\max}(c')$ points since otherwise there must be another
candidate that beats~$c$. Now, we show the tightness. The total number of votes is
$$|V_1^p|+ |V_2^p| +|V_3^p| = |\mathcal{S}|-|E|/3 +|\mathcal{S}|-|E|/3 + 3 |\mathcal{S}| = 5|\mathcal{S}|-2/3|E|.$$
In~$V_2^p$ and~$V_3^p$, candidate~$y$ is already fixed at the first
position in $2|\mathcal{S}|-1/3|E|$ votes and since in every vote 3
points have to be given, there are $3 \cdot (5|\mathcal{S}|-2/3|E|) -
2 \cdot (2|\mathcal{S}|-1/3|E|) = 11|\mathcal{S}|-4/3|E|$ points for the
remaining candidates left. The sum of  the maximum partial scores from all candidates from~$S \cup T \cup D \cup X \cup E$ is 
$$ 3\cdot 2 \cdot |\mathcal{S}| + |\mathcal{S}|-|E|/3 + 2 |\mathcal{S}| + 2|\mathcal{S}| -|E| = 11|\mathcal{S}|-4/3|E|.$$
To see this, note that clearly~$\sum_{e \in E}n_{e,3}= |\mathcal S|$ and $\sum_{e \in E}n_{e}= 3|\mathcal S|$. 
Thus, the tightness follows.

Now, we finally show the correctness of Claim~1. Due to the
tightness, the $|\mathcal{S}|-|E|/3$ candidates from~$X$ must take
position~2 in $|\mathcal{S}|-|E|/3$ votes from~$G^3$. Thus, there
remain $|E|/3$ second positions in~$G^3$ that are not fixed. Note that due to
tightness,  a candidate~$e_{i3}$ cannot take the third position
in~$v_i^3$. Hence, if the remaining second positions are not taken by
candidates from~$E$, we shift less than~$|E|/3$ candidates in~$G^3$, a
contradiction to Observation~1. Hence, these positions must be taken by
candidates from~$E$ and thus all second positions within~$G^3$ are
fixed. This implies that every candidate~$t_i$ from~$T$ must take either the
first or the third position in~$v_i^3$. More precisely, since $|E|/3$
candidates from~$E$ take a second position there must be $|E|/3$
candidates from~$T$ that take the first positions within the
corresponding votes. However, a candidate from~$T$ can only take the
first position if it makes zero points in $V_1^p\cup V_2^p$. Hence, there
must be $|E|/3$ candidates from~$T$, denoted as~$T'$, that make zero
points in $V_1^p\cup V_2^p$ and, due to tightness, all remaining candidates
from~$T$ must make 2 points in $V_1^p\cup V_2^p$. A candidate~$t_i \in T$
can make at most one point in~$V_1^p$ since due to the condition ``$s_i
\succ t_i$'' it shifts~$s_i$ to the first position (and
$s_p^{\max}(s_i) =2$). Hence, making two points within $V_1^p \cup V_2^p$ implies
that $t_i$ must make one point in $V_1^p$ and one point in~$V_2^p$ and
that the corresponding~$s_i$ must make 2 points in~$V_1^p$. This fixes
all positions in~$V_1^p \cup V_2^p$ and since a candidate~$s_i$ with $t_i
\in T'$ clearly makes zero points in~$V_1^p \cup V_2^p$, the correctness
of Claim~1 follows.  Altogether, we have that $\{S_i \mid t_i
\in T'\}$ forms an exact 3-cover for $(E,\mathcal{S})$.
\qed
\end{proof}

\subsection{Scoring vectors with $\al_1> \al_2= \dots = \al_{m-1}> 0$}\label{sec:3equal} 

In this subsection, we consider scoring rules defined by scoring
vectors that fulfill~$\al_1> \al_2= \dots = \al_{m-1}> 0$. Although
quite special, these rules might be of interest of their own. They can
be considered as a direct combination of the very common plurality and
veto rules where one allows to weight the contribution of the
plurality or veto part. For example, by using~$(10,1,\dots,1,0)$ the
``plurality'' part would have more influence to the outcome, whereas
for~$(10,9,\dots,9,0)$ the ``veto'' part would be more important. To
show NP-hardness, we give two types of many-one reductions from X3C;
one for the case~$\al_1<2 \cdot \al_2$ and one for the case~$\al_1 >
2\cdot \al_2$. As mentioned before, the case~$\al_1 = 2 \cdot \al_2$
remains open. Intuitively, for all other cases we make use of the
``asymmetry'' of the differences of the score values, that is, by
shifting a candidate from the first to the second position one
decreases its score by a different amount than by shifting it from the
last but one to the last position.  In the two following proofs,  the position in a linear order in which a candidate
gets~$\al_1$ points is denoted as
\emph{top position}, a position in which a candidate gets~$\al_2$
points as \emph{middle position}, and the position in which a
candidate gets zero points as
\emph{last position}.

\begin{theorem}An X3C-instance~$I$ can be reduced to a \textsc{Possible Winner}-instance for a scoring rule which produces a size-$m$ scoring vector satisfying the conditions $\al_1 > \al_2 = \al_{m-1} > \al_{m}=0$ and $\al_1 < 2 \cdot \al_2$ for $m= f(I) +2$. A suitable poly-type function~$f$ can be computed in polynomial time.\label{theo:3scoreI}
\end{theorem}

\begin{proof}
Let~$(E,\mathcal{S})$ denote an X3C-instance. We construct a partial profile~$P$ for which the
distinguished candidate~$c \in C$ is a possible winner if and only if $(E,\mathcal{S})$ is a yes-instance. The set of candidates is $C := \{c,h\} \uplus \{s_i,d_i,t_i \mid S_i \in \mathcal S\} \uplus  E$. The partial profile~$P$ consists of a set of partial
votes~$V^p$ and a set of linear orders~$V^l$. For~$1 \leq i \leq |\mathcal{S}|$, let~$S_i = \{e_{i1},e_{i2},e_{i3}\}$. Then the set of partial
votes~$V^p := V^p_1 \cup V^p_2$ is given by the following
subsets. The set~$V^p_1$ consists of $|E|/3$ copies of the partial vote
$$h \succ C \setminus \{h, s_1, \dots, s_{|\mathcal{S}|}\} \succ \{s_1, \dots, s_{|\mathcal{S}|}\}.$$
For every $i \in \{1, \dots, |\mathcal S|\}$, the set~$V^p_2$ contains the three votes

$v_i^1: h \succ C \backslash \{h,s_i,d_i\} \succ \{s_i,d_i\} $,

$v_i^2: e_{i1} \succ C \backslash \{e_{i1},t_i,d_i\} \succ t_i$, and

$v_i^3:  e_{i2} \succ C \backslash \{e_{i2}, e_{i3}, t_i\} \succ e_{i3}.$

Now, we pass on to the definitions of the maximum partial scores. To this
end, for a candidate~$e$ corresponding to an element~$e \in E$ (referred to as element candidate),
let~$n_{e,1+2}$ denote the number of subsets from~$\mathcal{S}$ in
which~$e$ is identical with~$e_{i1}$ or $e_{i2}$. Due to
Lemma~\ref{lem:linearvotes}, we can cast the linear votes such that
the following hold:

\begin{itemize}
\item $s_p^{\max}(s_i) = (|V^p|-1) \cdot \al_2$,
\item $s_p^{\max}(d_i) = s_p^{\max}(t_i) = \al_1 + (|V^p|-2) \cdot \al_2 $,
\item $s_p^{\max}(e) = (|V^p|- n_{e,1+2} +1) \cdot \al_2 +  (n_{e,1+2}-1) \cdot \al_1$,
\item $h$ is beaten by~$c$ in every extension.
\end{itemize}

The maximum partial scores of the element candidates are set such that
every element candidate has to be ``shifted'' to the right at least once. More
precisely, if a candidate~$e$ took the first position in all
votes in which it is identical with~$e_{i1}$ or~$e_{i2}$ and the
second position in all remaining votes (including the votes in which
it is identical with~$e_{i3}$), then~$s(e) = (|V^p|-n_{e,1+2}) \cdot
\al_2 + n_{e,1+2} \cdot \al_1 > s_p^{\max}(e)$
since~$\al_1>\al_2$. However, if, for any~$i$, $t_i$ or $d_i$ are
inserted at the first position in one of the votes in which~$e$
appears, then~$e$ makes at least~$\al_1-\al_2$ points less and thus is
beaten by~$c$. We denote this as
\textbf{Observation~2}. Now, we show the correctness of the construction.

\smallskip
\noindent
\textit{Claim:} Candidate~$c$ is a possible winner in~$P$ if and only if $(E,\mathcal{S})$ is a yes-instance.

\begin{figure}
$
\begin{array}{llllrll}
V_1^p:&       & h > &\dots& > s_i  &S_i \in Q\\
V_2^p:& v_i^1 & h > &\dots& > s_i >d_i & S_i \in Q\\
      & v_i^2 & d_i > e_{i1}>&\dots& > t_i & S_i \in Q\\
      & v_i^3 & t_i > e_{i2} >& \dots& > e_{i3} & S_i \in Q\\
\rowcolor{yellow}  & v_i^1 & h > &\dots& > d_i >s_i & \cellcolor{yellow} S_i \notin Q\\
\rowcolor{yellow}  & v_i^2 & e_{i1}> &\dots &> t_i >d_i & \cellcolor{yellow} S_i \notin Q\\
\rowcolor{yellow}  & v_i^3 & e_{i2}> &\dots &> e_{i3} > t_i &\cellcolor{yellow}  S_i \notin Q\\
\end{array}
$
\caption{Extension for the case~$\al_1>\al_2=\al_{m-1}>0$ and $\al_1 < 2 \cdot \al_2$. Extensions for candidates that do not correspond to subsets belonging to the solution set~$Q$ are highlighted.} \label{fig:kleiner2}
\end{figure}

\smallskip\noindent
``$\Leftarrow$'': Let~$Q$ denote an exact 3-cover
for~$(E,\mathcal{S})$. Then extend~$P$ as displayed in
Figure~\ref{fig:kleiner2}. More precisely, within~$V_1^p$ every
candidate~$s_i$ with~$S_i \in Q$ takes the last position in exactly
one of the $|E|/3$ votes.  Then, the candidates make the following
points within the extension of the partial votes. Every~$s_i$ takes the 
last position in one vote and middle positions in all other votes and thus
makes exactly~$s_p^{\max}(s_i)$ points.  For~$S_i \in Q$, every
candidate~$t_i$ and every candidate~$d_i$ takes one first and one last position, and
thus,~$s(d_i) = s(t_i)=\al_1 + (|V^p|-2)\cdot \al_2 =
s_p^{\max}(d_i) = s_p^{\max}(t_i)$. In the corresponding votes every element
candidate is shifted once since~$Q$ is an exact 3-cover and thus is
beaten by~$c$ due to Observation~$2$. Clearly, for~$S_i \notin Q$, $s_i$ is beaten by~$c$ as well. It remains to consider $d_i$ and $t_i$
with $S_i \notin Q$. Here, one has $s(d_i) = (|V^p|-1) \cdot \al_2 <
s_p^{\max}(d_i)$ and $s(t_i) = (|V^p|-1) \cdot \al_2 <
s_p^{\max}(t_i)$. Hence,~$c$ beats all other candidates and wins.

\smallskip\noindent
``$\Rightarrow$'': Consider an extension in which~$c$ wins. Due
to~$s_p^{\max}(s_i)$, every candidate~$s_i$ must take the last position
in at least one of the votes. Since $|V^p_1|= |E|/3$, at most $|E|/3$
candidates can take a last position in~$V_1^p$; denote the set of them
by~$S'$. Hence at least $|\mathcal{S}|- |E|/3$ candidates~$s_i$ must
take the last position in~$v_i^1$. Now, we show that for these
candidates the corresponding element candidates cannot be shifted to the right 
in~$v_i^2$ or $v_i^3$. Since~$s_i$ takes the last position in~$v_i^1$,
$d_i$ already makes $(|V^p|-1)\cdot \al_2$ in the extended partial
votes without~$v_i^2$. Hence,~$d_i$ must take the last position in
$v_i^2$ since otherwise~$s(d_i)= |V^p| \cdot \al_2 >
s_p^{\max}(d_i)$ because~$\al_1 < 2\al_2$. This implies that~$e_{i1}$ is not shifted and
that~$t_i$ takes a middle position in~$v_i^2$. Now, for~$t_i$ it
follows analogously that~$t_i$ must take the last position in~$v_i^3$
and thus neither~$e_{i2}$ nor~$e_{i3}$ is shifted. Altogether, this
means that all element candidates must be shifted by candidates
from~$S'$. Every~$s_i \in S'$ can shift three candidates by
setting~$s_i$ in the last position in~$v_i^1$ and $d_i$ and $t_i$ to
the first positions in $v_i^2$ and $v_i^3$, respectively. Since there
are $|E|$ element candidates, it follows that~$|S'| = |E|/3$ and that
all~$s_i\in S'$ must shift disjoint sets of element
candidates. Hence,~$S'$ corresponds to an exact 3-cover
for~$(E,\mathcal{S})$.
\qed
\end{proof}

In the remainder of this subsection, we consider the case that $\al_1
> 2 \cdot \al_2$. We also give a reduction from X3C. Note that
 the previous proof cannot be transferred directly and
thus we give a modified construction for which it will be more laborious
to show the correctness.

\begin{theorem}An X3C-instance~$I$ can be reduced to a \textsc{Possible Winner}-instance for a scoring rule which produces a size-$m$-scoring vector satisfying the conditions $\al_1 > \al_2 = \al_{m-1} > \al_{m}=0$ and $\al_1 > 2 \cdot \al_2$ for $m= f(I) +2$. A suitable poly-type function~$f$  can be computed in polynomial time.\label{theo:3score}
\end{theorem}

\begin{proof}
Let~$(E,\mathcal{S})$ denote an X3C-instance. Let~$k$ denote the size of a solution
for~$(E,\mathcal{S})$, that is, $k:= |E|/3$, and $t:= |\mathcal{S}|$.
 We construct a partial
profile~$P$ for which the distinguished candidate~$c \in C$ is a
possible winner if and only if $(E,\mathcal{S})$ is a
yes-instance. The set of candidates is~$C:= S
\uplus D \uplus E \uplus \{c,h\}$ with $S := \{s_i \mid 1 \leq i \leq
t\}$ (the subset candidates) and $D := \{d_i \mid 1 \leq i \leq t\}$, and $E$ (the element candidates).

Very roughly, the basic idea of the reduction is as follows. There are
three subsets of partial votes, in the first subset~$V_1^p$ one
``selects'' $t-k$ subset candidates from~$S$ that do not correspond to
an exact 3-cover and in the second subset~$V_2^p$ one selects $k$
subset candidates that correspond to an exact 3-cover. Selecting
hereby means that a solution subset candidate gets zero points in one
vote of~$V_2^p$ whereas every non-solution candidate gets~$\al_1$
points in a vote of~$V_1^p$. Hence, a solution candidate can make more
points than a non-solution candidate in the third subset~$V_3^p$.
Thus, a solution candidate can take a top position in~$V_3^p$ which
yields a cascading effect that makes it possible to shift the
corresponding element candidates such that they do not beat the
distinguished candidate~$c$.

Formally, the partial profile~$P$ consists of a set of partial
votes~$V^p$ and a set of linear orders~$V^l$. For~$1 \leq i \leq t$, let~$S_i = \{e_{i1},e_{i2},e_{i3}\}$, then the set of partial
votes~$V^p := V^p_1 \cup V^p_2 \cup V^p_3$ is given by the following
subsets.

\smallskip\noindent
\begin{tabular}{llllll}
 $V^p_1$:& $t-k$ copies of the partial vote &&$S \succ C\backslash(S\cup \{h\}) \succ h$\\
 $V^p_2$:& $k$ copies of the partial vote& &$h \succ C\backslash(S \cup \{h\}) \succ S$ \\
 $V^p_3$:& for  $1 \leq i \leq t$ the three partial votes&
$w_1^i$:&$ d_i \succ C\backslash \{d_i,e_{i1}, s_i\} \succ e_{i1}$  \\
&&$w_2^i$:& $h \succ  C\backslash \{d_i,e_{i2}, h\} \succ \{e_{i2},d_i\}$  \\
&&$w_3^i:$& $h \succ  C\backslash \{d_i,e_{i3}, h \} \succ \{e_{i3},d_i\}$\\
\end{tabular}

Note that in~$w_1^i$, candidate~$s_i$ can be inserted at any position.
The distinguished candidate~$c$ makes~$\al_2$ points in every partial
vote from~$V^p$. Hence, according to Lemma~\ref{lem:linearvotes}, we can set the linear orders  of~$V^l$ such that the following holds. For~$i = 1, \dots, t$,  $$s^{\max}_p(s_i)  = (|V^p|-2) \cdot \al_2 + \al_1,$$
$$s^{\max}_p(d_i) = (|V^p|-2)\cdot \al_2 + \al_1 - z$$
with~$z = \al_1 \mod \al_2$ if $\al_1 < 3 \al_2$, and~$z = \al_2$, otherwise\footnote{Note that this maximum partial score does not exactly fulfill the conditions of Lemma~\ref{lem:linearvotes} if $z\not= \al_2$. However, the construction can be easily extended to work for this case as well. More precisely, in this case~$z= \al_1 - \lfloor \al_1/\al_2\rfloor \cdot \al_2$ and $\lfloor \al_1/\al_2\rfloor \leq 3$. Thus, in the construction given in the proof of Lemma~\ref{lem:linearvotes} one can add~$\al_1$ and ``subtract'' $\al_2$ as often as required. The subtraction can be accomplished by changing the role of the dummy~``$d$'' and~$d_i$ within  a block.}. Note that it holds that~$\al_2 \geq z$ and 
\begin{align}
\al_1 -z \geq 2\al_2. \label{zz}
\end{align}  
For all~$e \in E$,
$s^{\max}_p(e) = (|V^p|-1) \cdot
\al_2$, that is, $e$ must have the last position in one of the partial
votes. And,~$s^{\max}_p(h) \geq |V^p| \cdot \al_1$, that is, $h$ can
beat~$c$ in no extension.

We now prove the following claim.

\begin{figure}
\setlength{\arraycolsep}{.3cm}
$
\begin{array}{llllll}
V^p_1: && \cellcolor{yellow} s_i > C\backslash\{s_i,h\} > h& \cellcolor{yellow} \forall s_i\text{ with } S_i \notin \mathcal{S'}\\
V^p_2: && h > C\backslash\{s_i,h\} > s_i &  \forall s_i\text{ with } S_i \in \mathcal{S'}\\
V_3^p: & w_1^i &\cellcolor{yellow} d_i > C\backslash\{s_i,d_i\} > s_i& \cellcolor{yellow} \forall s_i\text{ with } S_i \notin \mathcal{S'}\\
&w_2^i& \cellcolor{yellow} h > C\backslash\{d_i,h\} > d_i& \cellcolor{yellow} \forall s_i\text{ with } S_i \notin \mathcal{S'}\\
&w_3^i& \cellcolor{yellow} h > C\backslash\{d_i,h\} > d_i& \cellcolor{yellow} \forall s_i\text{ with } S_i \notin \mathcal{S'}\\
&w_1^i & s_i > C\backslash\{s_i,e_{i1}\} > e_{i1} &  \forall s_i\text{ with } S_i \in \mathcal{S'}\\
&w_2^i  & h > C\backslash\{e_{i2},h\} > e_{i2} &  \forall s_i\text{ with } S_i \in \mathcal{S'}\\
&w_3^i  & h > C\backslash\{e_{i3},h\} > e_{i3} &  \forall s_i\text{ with } S_i \in \mathcal{S'}\\
\end{array}$

\caption{Extension of~$V^p$ for an exact 3-cover~$\mathcal{S'} \subseteq \mathcal{S}$. The middle positions are not given explicitly since the order of the candidates is irrelevant. Extensions for candidates which do not belong to the solution set~$\mathcal{S'}$ are highlighted.}
\label{fig:ext}
\end{figure}

\smallskip
\noindent
\textit{Claim:} Candidate~$c$ is a possible winner of~$(V,C)$ if and only if $(E,\mathcal{S})$ is a yes-instance for X3C.

\smallskip\noindent
``$\Leftarrow$'': Let~$S' \subseteq \mathcal{S}$ denote an
exact 3-cover for $(E,\mathcal{S})$. Then, we extend the partial
profile as follows (Figure~\ref{fig:ext}). If~$S_i \in S'$, then $s_i$
is placed at the last position in one vote of~$V_2^p$ and at a
 middle position in all other votes from~$V^p_1 \cup V^p_2$.
If~$S_i \notin S'$, then $s_i$ is placed at the first position in one
of the votes in~$V^p_1$ and at a middle position in all other
votes from~$V^p_1
\cup V_2^p$. This is possible since there are~$t-k$ top position
and~$k$ last positions that can be taken by candidates from~$S$
in~$V^p_1 \cup V^p_2$. In~$V_3^p$, every candidate~$s_i$ with~$S_i \in
\mathcal{S}'$ is placed at the top position and the corresponding
element candidates~$e_{i2},e_{i3}$ at the last position in the
respective votes. Every candidate~$s_i$ with~$S_i \notin
\mathcal{S}'$ is placed at the last position and the corresponding
element candidates~$e_{i2},e_{i3}$ are placed at a middle position.

In the described extension, the candidates make the following points
in~$V^p$.  Every candidate $s_i \in S$ takes exactly one top
position and exactly one last position in~$V^p$. 
 Hence~$s(s_i)
=s^{\max}_p(s_i)$. For the candidates of~$D$ one has to distinguish two
cases. First, if~$S_i \notin S$, then, $s(d_i) = (|V^p|-3) \cdot
\al_2 + \al_1 \leq s_p^{\max}(d_i)$ since~$\al_2 \geq z$. Second, if~$S_i \in S$, then $s(d_i) = |V^p| \cdot \al_2 = (|V^p|-2) \cdot \al_2 + 2\al_2\leq
(|V^p|-2) \cdot \al_2 + \al_1 -z = s_p^{\max}(d_i)$ because of
Inequality~(\ref{zz}). Finally, we have to consider the candidates
from~$E$. Since for every~$S_i$ in the 3-cover, the corresponding
element candidates $e_{i1},e_{i2}$, and~$e_{i3}$ get at the last
position, every candidate of~$E$ takes one last and $|V^p|-1$ middle
positions and thus makes~$(|V^p|-1) \cdot \al_2$ points. It
follows that~$c$  wins in the considered extension.

\smallskip\noindent
``$\Rightarrow$'': In an extension of~$V$ in which~$c$ is the winner,
every element candidate from~$E$ must take the last position in one
vote of~$V^p$. This is only possible in~$V_3^p$ since every element
candidate is already fixed at a middle position in~$V_1^p \cup V_2^p$. More precisely, for
every~$i$,~$e_{i1}$ gets a last position if~$s_i$ is inserted at a
middle or the top position in the corresponding vote~$w_1^i$
and~$e_{i2}$/$e_{i3}$ can get a last position only if~$d_i$ takes a
middle position in the corresponding vote~$w_2^i$/$w_3^i$.

To find out what this means for the other candidates, we have to go into details here. For~$i=1,\dots,t$, let~$b_i$ denote the ``benefit'', i.e., the maximum number of element candidates that can be put at a last position in~$V_3^p$ depending on where~$s_i$ is placed in~$w_1^i$. Then, we can show the following.

\smallskip\noindent
\textbf{Observation~3:}
\begin{enumerate}
 \item $b_i = 3$ if~$s_i$ is placed in a top position in~$w_1^i$.
  \item $b_i = 1$ if~$s_i$ is placed in a middle position in~$w_1^i$. 
\item $b_i = 0$ if~$s_i$ is placed in a last position in~$w_1^i$.
\end{enumerate}

To see Observation~3, note that if~$s_i$ is on the top position
in~$w_1^i$, then~$d_i$ can take the middle position in~$w_2^i$
or~$w_3^i$ since the corresponding score~$s(d_i) = |V^p|\cdot \al_2
\leq s_p^{\max}(d_i)$. Thus, all three element candidates can be shifted to the last position. If~$s_i$ is not placed on the top, but in the middle position, then~$e_{i1}$ is still shifted to the last position, but~$d_i$ must take the last position in~$w_2^i$ or~$w_3^i$ and thus neither~$e_{i2}$ nor~$e_{i3}$ can have a last position in~$w_2^i$ or~$w_3^i$. To see this, assume that~$d_i$ has the top position in~$w_1^i$ and a middle position in~$w_2^i$ or~$w_3^i$, then 

$\begin{array}{l}
s(d_i) \geq |V_1^p \cup V_2^p| \cdot \al_2 + (|V_3^p|-2) \cdot \al_2 + \al_1 = (|V^p|-2) \cdot \al_2 +
\al_1 > s_p^{\max}(d_i),
\end{array}$
 a contradiction.
If~$s_i$ is placed on the last position in~$w_1^i$, then $e_{i1}$ cannot take the last position in $V_3^p$, and  neither can~$e_{i2}$ and~$e_{i3}$, because~$d_i$ takes the first position in~$w_1^i$ and gets~$\al_1$ points and has to take the last position in both~$w_2^i$ and~$w_3^i$ by the same argument as before. 

In the following, we show that in an extension in which~$c$ wins,
in~$V_1^p$ there must be $t-k$ different subset candidates~$s_i$ that
take the top position and each of the remaining~$k$ (solution)
candidates of~$S$ must take one last position in~$V_2^p$. It directly follows by Observation~3 that for all non-solution candidates we must have
that~$b_i=0$ and thus every solution candidate must shift the three
corresponding element candidates that must be different from the
element candidates corresponding to  the other solution candidates.

For every~$i$, let~$t_i$ denote the number of top positions that~$s_i$ takes
within~$V_1^p$ and~$l_i$ the number of last positions that~$s_i$ takes
within~$V_2^p$.  Observe that the following conditions must hold.
\begin{eqnarray} \label{eq:li}
& & \sum_{i=1}^t l_i = k, \nonumber \\ 
& & \sum_{i=1}^t t_i = t-k, \mbox{ since every position must be taken,}\\
& & \sum_{i=1}^t b_i \geq 3k,  \mbox{ since there are $3k$ element candidates and each} \nonumber \\[-4mm]
& & \mbox{\hspace{1.6cm} one must take at least one last position}. \nonumber
\end{eqnarray}

In the following, our strategy consists of three steps:
\begin{itemize}
\item We first investigate the dependencies of $l_i,t_i$, and~$b_i$ upon each other. For that sake, we distinguish the cases
$l_i=0$, $l_i=1$, and $l_i\geq 2$.  
\item Second, based on these case
distinctions, we can show that the case~$l_i\geq 2$ is not possible,
that is, every~$s_i$ can have at most one last position
in~$V_2^i$. This will need the most technical effort and will directly imply~$t_i\leq 1$ for
all~$i$. 
\item Third, we show that there is no candidate~$s_i$ with~$l_i =
t_i = 1$, which will imply that only candidates with~$l_i = 1$
contribute with a positive benefit and can place their element
candidates at a last position. Since there are only~$k$ such candidates, they must correspond to an exact 3-cover.
\end{itemize}

\textbf{First step.} We show some dependencies of~$l_i,t_i$, and~$b_i$ by
systematically enumerating all possible cases. (In the argumentation that follows the case distinction we are only interested in upper bounds of~$b_i$. Hence, we omit to show lower bounds.)

\smallskip
\noindent
\begin{tabular}{ll}
{\bf Case I:} $l_i = 0$ &\textbf{a)} if~$t_i = 0$, then~$b_i \leq 1$,\\
&\textbf{b)} if $t_i=1$, then~$b_i = 0$,\\
&\textbf{c)} $t_i \geq 2$ is not possible.\\
\end{tabular}

\smallskip
\noindent
Proof of Case~I:\\ 
\text{Ia}) ($l_i= t_i = 0$): Assume~$b_i= 3$, i.e., $s_i$ is on the top position in~$w_1^i$ due to Observation~3. Then $s(s_i) = (|V^p|-1)\al_2 +  \al_1 > s_p^{\max}(s_i)$, a contradiction, hence~$b_i \leq 1$.\\
\text{Ib}) ($l_i= 0, \, t_i= 1$): Assume~$b_i = 1$, i.e., $s_i$ is on a middle position in~$w_1^i$ due to Observation~3. Then $s(s_i) = (|V^p|-1)\al_2 +  \al_1 > s_p^{\max}(s_i)$, a contradiction, hence~$b_i = 0$.\\
\text{Ic}) ($l_i= 0, \, t_i \geq 2$): Assume~$s_i$ takes the last position in~$w_1^i$, that is,~$s_i$ makes as few points as possible within this case. Then,
\begin{eqnarray*}
 s(s_i) & = & (|V^p|-t_i -1)\al_2 +  t_i\al_1 \\
& > & (|V^p| -t_i -1 +2(t_i-1))\al_2 + \al_1 \\
& > & s_p^{\max}(s_i),
\end{eqnarray*}
 a contradiction, hence this case is not possible.

\smallskip
\noindent
\begin{tabular}{ll}
{\bf Case~II: $l_i = 1$} &\textbf{a)} if~$t_i = 0$, then~$b_i \leq 3$,\\
&\textbf{b)}		 if $t_i=1$, then~$b_i \leq 1$,\\
&\textbf{c)}			 $t_i \geq 2$ is not possible.\\
\end{tabular}

\smallskip
\noindent
Proof of Case~II: \\
\text{IIa}) ($l_i= 1, \, t_i = 0$), trivial upper bound.\\ 
\text{IIb}) ($l_i= t_i = 1$) Assume~$b_i = 3$, i.e., $s_i$ is on the top position in~$w_1^i$ due to Observation~3. Then $s(s_i) = (|V^p|-3)\al_2 +  2\al_1 > s_p^{\max}(s_i)$, a contradiction, hence~$b_i \leq 1$.\\
\text{IIc}) ($l_i= 1, \, t_i \geq 2$): Even if  $s_i$ takes the last position in~$w_1^i$ one has  
\begin{eqnarray*}
 s(s_i) & = & (|V^p|-t_i -2)\al_2 +  t_i\al_1 \\
& > & (|V^p| -t_i -2 +2(t_i-1))\al_2 + \al_1 \\
& = & (|V^p| + t_i -4)\al_2 + \al_1 \\
& \geq & s_p^{\max}(s_i),
\end{eqnarray*}
 a contradiction, hence this case is not possible.

\smallskip
\noindent
\begin{tabular}{ll}
{\bf Case III: $l_i \geq 2$}& \textbf{a)} if~$t_i = l_i$, then~$b_i=0$,\\
& \textbf{b)}		 	 if $t_i=l_i -1$, then~$b_i \leq 1$,\\
& \textbf{c)}			if $t_i \leq l_i- 2$, then~$b_i\leq3$,\\
& \textbf{d)}  			 $t_i > l_i$ is not possible.\\ 
\end{tabular}

\smallskip
\noindent
Proof of Case~III:\\ 
\text{IIIa}) ($l_i \geq 2, \, t_i = l_i$): Assume~$b_i = 1$, i.e., $s_i$ is on a middle position in~$w_1^i$ due to Observation~3. Then 
\begin{eqnarray*}
 s(s_i) & = & (|V^p|-t_i -l_i)\al_2 +  t_i\al_1 \\
& = & (|V^p|- 2t_i)\al_2 +  t_i\al_1 \\
& > & (|V^p| - 2t_i +2(t_i-1))\al_2 + \al_1 \\
& = & (|V^p| -2 )\al_2 + \al_1 \\
& = & s_p^{\max}(s_i),
\end{eqnarray*}
 a contradiction, hence~$b_i = 0$.\\
\text{IIIb}) ($l_i \geq 2, t_i = l_i -1$): Assume~$b_i = 3$, i.e., $s_i$ is on the top position in~$w_1^i$ due to Observation~3, then 
\begin{eqnarray*}
 s(s_i) & = & (|V^p|-t_i -l_i - 1)\al_2 +  (t_i +1) \al_1 \\
& = & (|V^p|- 2t_i -2 )\al_2 +  (t_i +1)\al_1 \\
& > & (|V^p| - 2t_i -2 +2t_i)\al_2 + \al_1 \\
& = & (|V^p| -2 )\al_2 + \al_1 \\
& = & s_p^{\max}(s_i),
\end{eqnarray*}
a contradiction, hence~$b_i \leq 1$.\\
\text{IIIc})  ($l_i \geq 2, \, t_i \leq l_i -2$): trivial upper bound.\\
\text{IIId})  ($l_i \geq 2, t_i > l_i$): Then 
\begin{eqnarray*}
 s(s_i) & = & (|V^p|-t_i -l_i -1)\al_2 +  t_i\al_1 \\
& > & (|V^p| -t_i - l_i -1  +2(t_i-1))\al_2 + \al_1 \\
& = & (|V^p| + t_i -l_i -3)\al_2 + \al_1 \\
& \geq & s_p^{\max}(s_i),
\end{eqnarray*}
 a contradiction, hence this case is not possible.\\

\textbf{Second step.}  Using the previous case distinctions, we show that no subset candidate~$s_i$ can take more than one last position in~$V_2^p$. For this, without loss of generality, we assume that the candidates~$s_i$ are sorted in decreasing order according to their corresponding~$l_i$, i.e., 
$$\underbrace{s_1, \dots, s_j}_{l_i\geq 2},\underbrace{s_{j+1}, \dots, s_r}_{l_i = 1},\underbrace{s_{r+1}, \dots, s_t}_{l_i = 0}.$$

\noindent
\textit{Claim~1}: In an extension in which~$c$ wins, it holds that~$l_i\leq 1$ for all~$i$.

\smallskip\noindent
To prove  Claim~1, we show that~$j=0$. More specifically, we prove that~$j>0$ implies that the total benefit~$B:=\sum_{i=1}^t b_i$ is less than~$3k$. This means that not all~$3k$ element candidates can take a last position and thus~$c$ cannot win.

Assume that~$j>0$. We start to show how to distribute the last and the
first positions of~$V_1^p$ and~$V_2^p$ in order to maximize~$B$. For
that sake, let $T_j := \sum_{i = 1}^jt_i$ denote the number of top positions that were taken by the
first~$j$ candidates~$s_1, \dots, s_j$.  Now, we consider the remaining indices $i \in \{j+1, \dots,
t\}$. Since for all of them~$l_i \leq 1$, it must also hold~$t_i \leq
1$ (see Case~I and Case~II). Thus and because of  Equation~(\ref{eq:li}), there
must be at least $t-k-T_j$ candidates from $s_{j+1}, \dots, s_t$ with~$t_i =
1$. For both remaining cases ($l_i=1$ and~$l_i = 0$), the
benefit~$b_i$ is greater for the case~$t_i = 0$ than it is for the
case~$t_i = 1$ (cf. Case~I and Case~II). Hence, to maximize the total
benefit $B$, it is desirable to minimize the number of
candidates having~$t_i=1$. Since there are~$t-j$ indices greater
than~$j$ and $t_i$ must be equal to one for at least $t-k-T_j$ indices, there
are at most~$t-j -(t-k-T_j)=k+T_j-j$ indices with~$t_i=0$ (\textbf{Observation~4}).
Furthermore, 
for every index from~$\{j+1, \dots, s_r\}$,  by setting~$t_i$ to
zero or one, one can ``choose'' between~$b_i = 1$ and~$b_i= 3$
(Case~II). For the remaining indices, one can choose between~$b_i=0$
and~$b_i=1$ by setting~$t_i$ to zero or one (Case~I). We show by contradiction that choosing Case~IIa (which results in~$b_i=3$) as often as possible is the way to maximize~$B$:

Assume that Case~IIa holds, that is~$l_i =1$ and~$t_i=0$, is not chosen as
often as possible. Then, first, there must be an index~$i \in \{j+1,
\dots, r\}$ with~$t_i = 1$ and hence with~$b_i=1$ (Case~IIb). Second, there must be
an index~$x>r$ with~$t_x = 0$ and hence~$b_x=1$
(Case~Ia).
 Then setting~$t_i=1$
and~$t_x=0$ does not violate  Equation~(\ref{eq:li}) and has the following effect. 
\begin{itemize}
\item $b_i$ is increased by 2 (from 1 to 3),
 \item $b_x$ is decreased by 1 (from 1 to 0).
\end{itemize}
Thus, $B= \sum_{i=1}^t b_i$ was not maximal.

Now, we have argued that to maximize~$B$, one has to choose Case~IIa as often as possible (\textbf{Observation~5}). Using this, we can  compute the maximal value~$\max B$ of~$B$ (showing that is must be less than~$3k$). 
For that sake, we  first consider  the benefit coming from the first~$j$ candidates~$s_1, \dots, s_j$, which we denote by~$B_j := \sum_{i = 1}^j b_i$.
Let $B_j^0$ denote the set of indices $i \in \{1, \dots, j\}$ with $b_i=0$, let $B_j^1$ denote the set of indices $i \in \{1, \dots, j\}$ with $b_i=1$, and let $B_j^3$ denote the set of indices $i \in \{1, \dots, j\}$ with $b_i=3$.
 Then,  Case~III directly gives the following bound for the number of top positions assumed by the first~$j$ candidates.
\begin{equation}T_j \leq \sum_{i \in B_j^0}l_i + \sum_{i \in B_j^1}(l_i -1) + \sum_{i \in B_j^3} (l_i -2) = \sum_{i=1}^jl_i-|B_j^1|-2|B_j^3|,\label{eq:Tj}
\end{equation}
which will be needed in the following.

Due to the previous discussion we know that in the remaining
positions, we have to choose~$t_i=0$ for~$k+T_j-j$ indices (cf.~Observation~4) and one should choose Case~IIa, that is, $l_i=1$
and~$t_i=0$, as often as possible (cf.~Observation~5). Clearly,~$l_i=1$ must be chosen~$k-\sum_{i=1}^jl_i$
times whereas there are~$k+ T_j -j$ indices with~$t_i=0$. Hence, to
compute a total upper bound on~$B$, we have to distinguish two cases:
First, $k-\sum_{i=1}^jl_i \leq k+T_j-j$, and, second,
$k-\sum_{i=1}^jl_i > k+T_j-j$.

For the first case, we obtain
\begin{eqnarray*}
\max B & = & \underbrace{|B_j^1| + 3|B_j^3|}_{B_j} + 3 \underbrace{(k-\sum_{i=1}^j l_i)}_{l_i=1, \, t_i=0} + \underbrace{k+T_j -j - (k - \sum_{i=1}^j l_i)}_{l_i=0, \, t_i=0} \\
 & = & |B_j^1| + 3|B_j^3| + 3k -2\cdot \sum_{i=1}^j l_i + T_j-j\\
 & \stackrel{(\ref{eq:Tj})}{\leq} &  |B_j^1| + 3|B_j^3| + 3k -2\cdot \sum_{i=1}^j l_i + \sum_{i=1}^jl_i-|B_j^1|-2|B_j^3|-j\\
&= &3k -  \sum_{i=1}^j l_i -j + |B_j^3|
\end{eqnarray*}

Since~$|B_j^3| \leq j$ it holds that the maximal value of~$B$ is strictly less than~$3k$ for~$j\geq 1$. Thus, at least one element candidate does not take a last position and hence beats~$c$, a contradiction.  

For the second case,  we obtain
\begin{eqnarray*}
\max B & = & \underbrace{|B_j^1| + 3|B_j^3|}_{B_j} + 3 \underbrace{(k+T_j-j)}_{l_i=1, \, t_i=0} + \underbrace{k - \sum_{i=1}^j l_i -(k+T_j-j)}_{l_i=1, \, t_i=1} \\
& = & |B_j^1| +3|B_j^3| +3k +2T_j-2j-\sum_{i=1}^j l_i\\
& \stackrel{(\ref{eq:Tj})} \leq & |B_j^1| +3|B_j^3| +3k + \sum_{i=1}^j l_i -|B_j^1| -2|B_j^3| + T_j-2j-\sum_{i=1}^j l_i\\
& =& 3k +|B_j^3| +T_j -2j
\end{eqnarray*}
Furthermore, in this case it follows directly from $k-\sum_{i=1}^jl_i > k+T_j-j$ that~$\sum_{i=1}^j l_i +T_j <j$. For~$j>0$ this means that~$T_j <j$. By definition, we have~$|B_j^3| \leq j$, and thus $\max B$ is less than~$3k$. 
This completes the proof of  Claim~1. 
We therefore have~$j=0$ which means~$l_i \leq 1$ for all~$i \in \{1, \dots, t\}$ and thus also~$t_i\leq 1$ for all~$i$ (Case~I and~II). 

\textbf{Third step.} We now show that there cannot be any candidate~$s_i$ which takes one last position and one first position in~$V_1 \cup V_2$, i.e. we cannot have~$t_i=l_i=1$ for any~$s_i$. 
Assume that the set of candidates~$Q:=\{s_i\mid t_i = l_i = 1\}$ is not empty. Then, due to Observation~3, the maximum value of~$B$ is
\[\max B = \underbrace{1 \cdot |Q|}_{l_i = t_i = 1} + \underbrace{3\cdot(k - |Q|)}_{l_i = 1, t_i = 0} + \underbrace{0}_{l_i= 0, t_i = 1} + \underbrace{1\cdot|Q|}_{l_i=t_i=0} = 3k - |Q|, \] 
a contradiction. 
Thus,~$t-k$ many of the subset candidates~$s_i$ take a top position in~$V_1^p$, and the remaining~$k$ subset candidates take a last position in~$V_2^p$. Now, each of these~$k$ candidates must place its corresponding element candidates at the last positions in~$V_3^p$.
Since~$c$ can only be a winner if each of the $3k$ element candidates
takes a last position in a vote from~$V^p_3$ and in total at
most~$3k$ element candidates can take a last position in~$V^p_3$,
every element candidate must take exactly one last position. Thus,
for~$i\not = j$ such that~$s_i$ and~$s_j$ take a last position in~$V_2^p$,
$\{e_{i1},e_{i2},e_{i3}\}$ and $\{e_{j1},e_{j2},e_{j3}\}$ must be
disjoint. It follows that $\{S_i \mid s_i$ takes a last position
in~$V_2^p\}$ forms an exact 3-cover.
 \qed
\end{proof}

\section{Putting  all together}
\label{sec:main}
We are now ready to combine the many-one reductions from the previous sections
to one general reduction.  Basically, the problem we encounter by using one specific  reduction from the
previous sections is that such a reduction produces a
\textsc{Possible Winner}-instance with a certain number~$m$ of
candidates. Thus,  one needs to ensure that the
 size-$m$ scoring vector  provides a sufficient number
of positions with equal/different scores. This seems not to be
possible in general. However, for every specific instance of \textsc{Exact Cover By 3-Sets} or \textsc{Multicolored Clique}, we can compute a number of positions with equal or different
scores that is sufficient for the corresponding reduction, and we can use
the maximum of all these numbers for the combined reduction. This is
the underlying idea for the following proof.

\begin{theorem}
\textsc{Possible Winner} is NP-complete for a   scoring rule~$r$ if there is a constant~$z$ such that all scoring vectors produced by~$r$ for more than~$z$ candidates are different from $(0,\dots,0)$, $(1,0,\dots, 0)$, $(1,\dots,1,0)$, and $(2,1, \dots,1,0)$. 
\label{theo:maingeneral}
\end{theorem}

\begin{proof}
We give a reduction from X3C restricted to instances of size greater
than~$z$ to
\textsc{Possible Winner} for~$r$. 
Let~$I$ with $|I|>z$ denote an X3C-instance. Since X3C and MC are NP-complete, there is a polynomial-time
reduction from X3C to MC. Hence, let~$I'$ denote an MC-instance whose
size is polynomial in~$|I|$ and which is a yes-instance if and only
if~$I$ is a yes-instance.

Let $f_1$ denote a poly-type function  to compute the number of different score values  as stated for Theorem~\ref{theo:aaai}, $f_1'$ as for Theorem~\ref{theo:approval2}, $f_2'$ as for Lemma~\ref{lem:I}, $f_2$ as for Lemma~\ref{lem:II}, $f_3$ as for Lemma~\ref{lem:III}, $f_4$ as for Lemma~\ref{lem:IV}, and $f_5$ as for Theorem~\ref{theo:3score}. 
Define $x:= \max\{f_1(I), f'_1(I'), f'_2(I'), f_2(I), f_3(I), f_4(I), f_5(I)\}$ and consider the scoring vector~$\overrightarrow{\al}$ of
size~$x\cdot (x+1)$ produced by~$r$. Then we show the following.

\smallskip\noindent
\textit{Claim:}
 For~$\overrightarrow{\al}$ it holds  that $|\{ i \mid \al_i > \al_{i+1} \}| \geq x$ or  that~$\al_i = \dots = \al_{i+x}$ for some position~$i$.

The correctness of the claim can be seen as follows.  First, assume
that $\overrightarrow{\al}$ does not fulfill~$\al_i > \al_j$ for~$x$
different positions~$i$. Then consider $x \cdot (x+1)$ indices
of~$\overrightarrow{\al}$. Since they can have at most~$x$ different
score values, there must be a single score value that is assigned to at
least~$x+1$ indices, that is, there is an index~$i$ with~$\al_i= \dots
= \al_{i+x}$.  Second, if there is no index~$i$ such that~$\al_i =
\dots = \al_{i+x}$ for a position~$i$, then again consider~$x \cdot
(x+1)$ indices of~$\overrightarrow{\al}$. Since each score value can
be assumed at most~$x$ times, there must be at least~$x$ different
score values.

\begin{table}[t]
\caption{Subcases for scoring rules having an unbounded number of equal score values.}\label{tab:equalred}
\begin{tabular}{llll}
\toprule
Case~I \quad & $\exists i \leq m-1$ s.t. &  $\al_{i-x}= \dots = \al_{i-1} > \al_i$&  Theorem~\ref{theo:approval2}\\
Case~IIa\quad & $\exists i\geq 2, \, \exists j<i$ s.t. & $\al_{i} > \al_{i+1} = \dots = \al_{i+x}$ and $\al_{j} < 2 \al_{j+1}$&   Lemma~\ref{lem:I}\\
Case~IIb\quad& $\exists i \geq 2, \, \exists j<i$ s.t. & $\al_{i} > \al_{i+1} = \dots = \al_{i+x}$ and $\al_j \geq 3 \al_i$ &Lemma~\ref{lem:II}\\
Case~IIc \quad&& $(\al_1,\al_2,0,\dots, 0)$ and $3\al_2 >\al_1>2 \al_2$& Lemma~\ref{lem:III}\\
Case~IId\quad& & $(2,1,0,\dots, 0)$ & Lemma~\ref{lem:IV}\\
Case~III\quad& &  $\al_1 > \al_2 = \al_{m-1} > \al_{m}=0$ and $\al_1 \neq 2 \cdot \al_2$& Theorem~\ref{theo:3score}\\
\bottomrule
\end{tabular}

\end{table}

Now, due to the Claim, we can distinguish two main cases. If~$\overrightarrow{\al}$ has at
least $x$ different score values, then we apply the X3C-reduction
given in Theorem~\ref{theo:aaai}. Otherwise, we have an unbounded number of equal score values. In this case we distinguish the  subcases given in Table~\ref{tab:equalred}. 
For all these subcases, there are many-one reductions used in the corresponding lemmata/theorems. Hence, it remains to show that each scoring vector can be handled by at least one of these cases. Clearly, $\overrightarrow{\al}$ must  have the form $\al_{i-x}= \dots = \al_{i-1} > \al_i$ for an  $i \leq m-1$ (Case~I),  or $\al_{i} > \al_{i+1} = \dots = \al_{i+x}$ for $i \geq 2$ (Case~II), or $\al_1 > \al_2 = \al_{m-1} > \al_{m}=0$ and $\al_1 \neq 2 \cdot \al_2$~(Case~III). For Case~I and Case~III, the existence of many-one reductions follows immediately from the corresponding Theorems~\ref{theo:approval2} and~\ref{theo:3score}.  Thus, it remains to discuss Case~II, the case that  $\overrightarrow{\al}$ has the form $\al_{i} > \al_{i+1} = \dots = \al_{i+x}$ for $i \geq 2$.

To this end,  we start with the case~$i>2$. Clearly, there must be at least
three scoring values which are not equal to zero, namely,~$\al_{i-2},\al_{i-1}$,
and~$\al_i$. If one has~$\al_{i-1}< 2\al_i$ or~$\al_{i-2}< 2\al_{i-1}$, then NP-hardness follows directly from
Lemma~\ref{lem:I}. Otherwise, one must have~$\al_{i-1} \geq 2\al_{i}$ and~$\al_{i-2}\geq 2\al_{i-1}$. Hence, it follows
that~$\al_{i-2} \geq 4\al_i$ and NP-hardness follows directly
from Lemma~\ref{lem:II}. It remains to consider all scoring rules of type~$(\al_1,\al_2,0,\dots, 0)$. Here, we can distinguish the following four cases:
\begin{itemize}
\item $\al_1< 2\al_2$: NP-hardness follows from Lemma~\ref{lem:I},
\item $\al_1 = 2 \al_2$: NP-hardness follows from Lemma~\ref{lem:IV},
\item $2\al_2<\al_1< 3\al_2$: NP-hardness follows from Lemma~\ref{lem:III}, and
\item $\al_1 \geq 3\al_2$: NP-hardness follows from Lemma~\ref{lem:II}.
\end{itemize}
Since the membership in NP is obvious, the main theorem follows.
\qed 
\end{proof}

\paragraph{Pure scoring rules}
Based on all previous considerations, for pure scoring rules we almost arrive at a dichotomy. More precisely, we can state the following.

\begin{theorem}\textsc{Possible Winner} is NP-complete for all non-trivial pure scoring rules 
except plurality, veto, and scoring rules for which there is a constant~$z$ such that the produced  scoring
vector is~$(2,1,\dots,1,0)$ for every number of candidates greater than~$z$. For 
plurality and veto it is solvable in polynomial time.
\label{theo:main}
\end{theorem}

\begin{proof}
Plurality and veto are polynomial-time solvable due to
Proposition~\ref{prop:poly}. Having any non-trivial scoring vector different from $(1,0,\dots,0)$, $(1,\dots,1,0)$, and $(2,1,\dots,1,0)$ for~$m$ candidates, it is not possible to obtain a scoring vector of one of these three types (or~$(0,\dots, 0)$) for $m'>m$ by inserting scoring values. Hence, since we only consider
pure scoring rules,  the scoring rule does not produce a scoring vector of type plurality, veto, $(0,\dots , 0)$, or
$(2,1,\dots,1,0)$ for all~$m \geq z$. Then the statement follows by Theorem~\ref{theo:maingeneral}.
\qed
\end{proof}

\paragraph{``Non-pure'' scoring rules}\label{sec:hybrid}
We end this section with a brief informal discussion about the problem
of classifying non-pure scoring rules in general. As stated in
Theorem~\ref{theo:maingeneral}, we can show NP-hardness for non-pure
scoring rules if (starting from a constant number of candidates) all produced scoring
vectors are ``difficult''. Clearly, it is possible to extend the range
of NP-hardness results to scoring rules that produce only few ``easy''
vectors; for example, having a difficult vector for all odd numbers of
candidates and an easy vector for all even ones.  However, this is not
possible in general. Roughly speaking, if the underlying difficult
part of the language becomes too sparse, then there cannot be a
many-one reduction from an NP-complete problem since the densities of
the problems are not polynomially related (see e.g.~\cite{Pap94}).
Note that this situation does not appear for the dichotomy result from
Hemaspaandra and Hemaspaandra~\cite{HH07} for \textsc{Manipulation}
for weighted voters. The intuitive reason for this is that their
reductions for the NP-hardness in the case of weighted voters already hold for a constant number of
candidates (and all scoring rules except plurality are NP-hard in this case).

\section{Conclusion and outlook}

In this work, we settled the computational complexity for
\textsc{Possible Winner} for almost all pure scoring rules. More
precisely, the only case that was left open regards the scoring rule
defined by the scoring vector~$(2,1,\dots,1,0)$, whereas for all other
rules except plurality and veto, we obtained NP-completeness
results. In a very recent work, Baumeister and Rothe~\cite{BR10}
completed the dichotomy by showing the NP-completeness of
\textsc{Possible Winner} for the case of~$(2,1, \dots, 1,0)$.

A natural next step of research is to investigate algorithmic
approaches that deal with NP-hard problems like approximation
algorithms or ``efficient'' exponential-time algorithms. Here, an
interesting approach is to consider the parameterized
complexity~\cite{DF99,FG06,Nie06} and its sequel multivariate algorithmics~\cite{Nie10}. There are  first considerations
for several voting rules~\cite{BHN09} as well as fixed-parameter
tractability results for \textsc{Possible Winner} for $k$-approval
with respect to the combined parameter ``number of partial votes''
and~$k$~\cite{Bet10}. A parameter of general interest is the
``number of candidates''. In this case,  \textsc{Possible Winner} is 
shown to be fixed-parameter tractable for several voting systems using
a powerful classification framework based on integer linear programming   but still lacks efficient combinatorial fixed-parameter algorithms~\cite{BHN09}.  Furthermore, multivariate complexity
analysis might offer a way to tackle the
\textsc{Possible Winner} problem for voting systems for which the
``normal'' winner determination is already NP-hard. For example, there
are recent studies for  Kemeny, Dodgson, and Young elections that
contain parameterized algorithms with respect to several 
parameters~\cite{BFGNR09,BGKN10,BGN09,Sim09}.
 It is open whether such 
results can be achieved for the \textsc{Possible Winner} problem.

The \textsc{Possible Winner} problem not only generalizes the
\textsc{Manipulation} problem but also comprises other relevant special
cases. For example, very recently, Chevaleyre et al.~\cite{CLMM10} investigated the
computational complexity of the following problem: Given a set of
linear votes, an integer $s$, and a distinguished candidate~$c$, can
one add $s$ candidates such that $c$ becomes a winner? There is reasonable hope to achieve more positive algorithmic results for this and other relevant special cases of \textsc{Possible Winner}.

A further direction of future research regards the counting version of
\textsc{Possible Winner}~\cite{BBF10}. Here, one wants to find out in how many
extensions a distinguished candidate wins. Answering this question
allows to compare two candidates that are possible winners.

\paragraph{Acknowledgments} We thank  Rolf Niedermeier and Johannes Uhlmann for fruitful discussion  and helpful comments.  
 We are very grateful to the anonymous referees of \textit{MFCS'09} and \textit{JCSS} for constructive and beneficial feedback that helped to significantly improve this work.

\bibliographystyle{abbrv}
\bibliography{posswin}

\end{document}

%% file: possflownad.pstex_t
\begin{picture}(0,0)%
\includegraphics{possflownad.pstex}%
\end{picture}%
\setlength{\unitlength}{4144sp}%
\begingroup\makeatletter\ifx\SetFigFont\undefined%
\gdef\SetFigFont#1#2#3#4#5{%
  \reset@font\fontsize{#1}{#2pt}%
  \fontfamily{#3}\fontseries{#4}\fontshape{#5}%
  \selectfont}%
\fi\endgroup%
\begin{picture}(8595,4456)(1711,-3782)
\put(4261,-1486){\makebox(0,0)[lb]{\smash{{\SetFigFont{29}{34.8}{\rmdefault}{\mddefault}{\updefault}{\color[rgb]{0,0,0}$v_4$}%
}}}}
\put(4441,314){\makebox(0,0)[lb]{\smash{{\SetFigFont{29}{34.8}{\rmdefault}{\mddefault}{\updefault}{\color[rgb]{0,0,0}$v_1$}%
}}}}
\put(7201,239){\makebox(0,0)[lb]{\smash{{\SetFigFont{29}{34.8}{\rmdefault}{\mddefault}{\updefault}{\color[rgb]{0,0,0}$a$}%
}}}}
\put(7216,-1531){\makebox(0,0)[lb]{\smash{{\SetFigFont{29}{34.8}{\rmdefault}{\mddefault}{\updefault}{\color[rgb]{0,0,0}$b$}%
}}}}
\put(7186,-3316){\makebox(0,0)[lb]{\smash{{\SetFigFont{29}{34.8}{\rmdefault}{\mddefault}{\updefault}{\color[rgb]{0,0,0}$d$}%
}}}}
\put(2881,-3016){\makebox(0,0)[lb]{\smash{{\SetFigFont{29}{34.8}{\rmdefault}{\bfdefault}{\updefault}{\color[rgb]{0,0,0}1}%
}}}}
\put(3046,-931){\makebox(0,0)[lb]{\smash{{\SetFigFont{29}{34.8}{\rmdefault}{\bfdefault}{\updefault}{\color[rgb]{0,0,0}1}%
}}}}
\put(5656,-541){\makebox(0,0)[lb]{\smash{{\SetFigFont{29}{34.8}{\rmdefault}{\bfdefault}{\updefault}{\color[rgb]{0,0,0}1}%
}}}}
\put(5776,134){\makebox(0,0)[lb]{\smash{{\SetFigFont{29}{34.8}{\rmdefault}{\bfdefault}{\updefault}{\color[rgb]{0,0,0}1}%
}}}}
\put(5791,-1741){\makebox(0,0)[lb]{\smash{{\SetFigFont{29}{34.8}{\rmdefault}{\bfdefault}{\updefault}{\color[rgb]{0,0,0}1}%
}}}}
\put(5386,-1171){\makebox(0,0)[lb]{\smash{{\SetFigFont{29}{34.8}{\rmdefault}{\bfdefault}{\updefault}{\color[rgb]{0,0,0}1}%
}}}}
\put(4936,-2551){\makebox(0,0)[lb]{\smash{{\SetFigFont{29}{34.8}{\rmdefault}{\bfdefault}{\updefault}{\color[rgb]{0,0,0}1}%
}}}}
\put(8326,-3211){\makebox(0,0)[lb]{\smash{{\SetFigFont{29}{34.8}{\rmdefault}{\mddefault}{\updefault}{\color[rgb]{0,0,0}$s(c) - 1$}%
}}}}
\put(5626,-3436){\makebox(0,0)[lb]{\smash{{\SetFigFont{29}{34.8}{\rmdefault}{\bfdefault}{\updefault}{\color[rgb]{0,0,0}1}%
}}}}
\put(3466,-1666){\makebox(0,0)[lb]{\smash{{\SetFigFont{29}{34.8}{\rmdefault}{\bfdefault}{\updefault}{\color[rgb]{0,0,0}1}%
}}}}
\put(1711,-1681){\makebox(0,0)[lb]{\smash{{\SetFigFont{29}{34.8}{\rmdefault}{\mddefault}{\updefault}{\color[rgb]{0,0,0}$s$}%
}}}}
\put(9676,-1771){\makebox(0,0)[lb]{\smash{{\SetFigFont{29}{34.8}{\rmdefault}{\mddefault}{\updefault}{\color[rgb]{0,0,0}$t$}%
}}}}
\put(4321,-3301){\makebox(0,0)[lb]{\smash{{\SetFigFont{29}{34.8}{\rmdefault}{\mddefault}{\updefault}{\color[rgb]{0,0,0}$v_5$}%
}}}}
\put(7606,-2266){\makebox(0,0)[lb]{\smash{{\SetFigFont{29}{34.8}{\rmdefault}{\mddefault}{\updefault}{\color[rgb]{0,0,0}$s(c)-1$}%
}}}}
\put(8236,-781){\makebox(0,0)[lb]{\smash{{\SetFigFont{29}{34.8}{\rmdefault}{\mddefault}{\updefault}{\color[rgb]{0,0,0}$s(c) - 1$}%
}}}}
\end{picture}%